\documentclass[11pt]{article}
\usepackage[dvips]{graphicx}
\usepackage{amssymb,amsmath}
\usepackage{graphics}
\usepackage{epsfig,amsfonts,amssymb}
\usepackage{txfonts}
\usepackage{mathrsfs}
\usepackage{cite}
\usepackage{mathtext}
\usepackage{epstopdf}
\usepackage{epsfig}
\usepackage{psfrag}
\usepackage{amscd}
\usepackage[english]{babel}
\usepackage[latin1]{inputenc}
\usepackage{times}

\begin{document}

\begin{center}
{\Large  An AdS/CFT Connection between Boltzmann and Einstein}\\
{Kinetic Theory and Pure Gravity in AdS}
\end{center}
\centerline{\large \rm Ramakrishnan Iyer$^{1}$, Ayan Mukhopadhyay$^{2}$}
\bigskip
\centerline{\large \it $^1$Department of Physics and Astronomy}

\centerline{\large \it University of Southern California
Los Angeles California 90089-0484, USA}
\vspace*{1.0ex}

\centerline{E-mail: ramaiyer@usc.edu}

\vspace*{5.0ex}

\centerline{\large \it $^2$Harish-Chandra Research Institute}

\centerline{\large \it  Chhatnag Road, Jhusi,
Allahabad 211019, INDIA}
\vspace*{1.0ex}

\centerline{E-mail: ayan@mri.ernet.in}

\begin{abstract}
The AdS/CFT correspondence defines a sector with universal strongly coupled dynamics in the field theory as the dual of pure gravity in AdS described by Einstein's equation with a negative cosmological constant. We explain here, from the field-theoretic viewpoint how the dynamics in this sector gets determined by the expectation value of the energy-momentum tensor \emph{alone}. We first show that the Boltzmann equation has very special solutions which could be \textit{functionally} completely determined in terms of the energy-momentum tensor alone. We call these solutions \textit{conservative solutions}. We indicate why conservative solutions should also exist when we refine this kinetic description to go closer to the exact microscopic theory or even move away from the regime of weak coupling so that no kinetic description could be employed. We argue that these \textit{conservative solutions} form the universal sector dual to pure gravity at strong coupling and large $N$. Based on this observation, we propose a \textit{regularity condition} on the energy-momentum tensor so that the dual solution in pure gravity has a smooth future horizon. We also study if irreversibility emerges only at long time scales of observation, unlike the case of the Boltzmann equation.

\end{abstract}

\newpage  \setcounter{footnote}{0}
\section{INTRODUCTION}
The AdS/CFT correspondence \cite{Maldacena} has opened up new vistas in understanding strongly coupled phenomena in four-dimensional conformal field theories. At strong coupling and large $N$, the dual classical theory of gravity admits a consistent truncation to pure gravity in asymptotically $AdS_{5}$ space and is described by Einstein's equation with a negative cosmological constant. Any such solution of Einstein's equation, which has a smooth future horizon, describes a dual state of the field theory at finite temperature and its dynamics. The final temperature of the horizon corresponds to the temperature of final equilibrium in the field theory. The dynamics of gravity is governed only by the five-dimensional Einstein's equation in this sector. The dynamics is therefore always universal, i.e. it is completely independent of the detailed matter content and couplings of the field theory. In fact any conformal gauge theory at strong 't Hooft coupling and large $N$, which has a gravity dual, must contain this universal sector \cite{footnote1}. This universal sector is practically important because it can describe, for instance, hydrodynamics of the dual conformal field theory (for a recent review, please see \cite{Rangamani}).

However this universal sector, as described by gravity, covers a variety of phenomena well beyond the hydrodynamic regime. Such phenomena also include decoherence\cite{footnote2}. It would then be certainly useful to understand the universal sector completely and also to decipher unambiguously the corresponding phenomena in the field theory side. This motivates our work.

In the universal sector the dynamics of all gauge fields and higher form fields have been turned off on the gravity side. This means that in the dual field theory all chemical potentials vanish and therefore all conserved charge currents are absent. We strictly restrict ourselves to the case of zero global angular momentum in the field theory configurations, so that the final equilibrium configuration in gravity is always a static black brane. We will not explicitly mention this restriction in the rest of this paper; certainly most of the results here can be readily generalized to the case of nonzero global angular momentum. We also limit ourselves to the case of a flat boundary metric, so that the gauge theory is living in (the conformal class of) flat Minkowski space. The boundary topology then is $R^{4}$ and the static black brane is the unique equilibrium configuration in the absence of all other conserved charges. Metastable configurations like small black holes do not appear with this choice of boundary topology and boundary metric.

The crucial aspect of the universal sector is that all the solutions in gravity are uniquely determined by the boundary metric and the boundary stress tensor. In particular, when the boundary metric is flat, we have good understanding \cite{0002230, Bjorken, 0810.4851} about the general features of the five-dimensional geometry \cite{footnote3}. It follows that all the states in the dual field theory constituting the universal sector are also uniquely specified by the expectation value of the energy-momentum tensor alone. The dynamics of these states can then be determined by simply following the evolution of the energy-momentum tensor alone. This necessitates an understanding, from the field-theoretic point of view, as to how all observables and their time evolution could be functionally determined by the energy-momentum tensor.

Here we will address this question; first in the regime of weak coupling, so that we can employ the quasiparticle description and also use kinetic theories, which are coarse-grained descriptions of microscopic laws. Specifically, we use the Boltzmann equation which has proven useful \cite{Arnold,RHIC} in determining the shear viscosity and higher order hydrodynamic transport coefficients and the relaxation time in weakly coupled gauge field theories. It has also been shown that an effective Boltzmann equation can be used to study nonequilibrium phenomena in high temperature QCD and is equivalent to an exact perturbative treatment \cite{Arnold}. Despite being a coarse-grained description, the Boltzmann equation retains the power to describe nonequilibrium phenomena far away from the hydrodynamic regime and at length scales and time scales shorter than the mean-free path and the relaxation time respectively. However it is not applicable to phenomena at microscopic length and time scales.

We prove that there exist very special solutions of the Boltzmann equation which are functionally determined by the energy-momentum tensor alone. We call such solutions ``conservative solutions''. These solutions, although very special, constitute phenomena far away from equilibrium and well beyond the hydrodynamic regime. The existence of conservative solutions can be conveniently proven for nonrelativistic monoatomic gases using some basic structural properties of the Boltzmann equation and can be easily extended to include relativistic and semiclassical corrections. We show that these solutions can be constructed even for multicomponent systems relevant for relativistic quantum gauge theories.

It will thus be natural to make the assumption that the conservative solutions constitute the universal sector of strongly coupled gauge theories with gravity duals. This will explain why the states in the universal sector are determinable functionally by the energy-momentum tensor alone. This assumption, through the AdS/CFT correspondence, will have powerful consequences for gravity. The same condition on the energy-momentum tensor,required to make the state in the field theory a conservative solution, will now be required to make the dual solution in gravity have a smooth future horizon. In other words, the \textit{conservative} condition on the energy-momentum tensor in field theory should now transform into the \textit{regularity} condition in gravity. We will use this observation (assumption) to propose five equations which, in combination with the four conservation (or hydrodynamic) equations, will provide a framework to determine the evolution of the energy-momentum tensor. Any energy-momentum tensor satisfying these equations will give us solutions in gravity with smooth future horizons. This framework has sufficient predictive power to determine nonequilibrium states in the universal sector beyond hydrodynamics in a systematic perturbative manner, given that the purely hydrodynamic energy-momentum tensor up to second order in the derivative expansion \cite{Bjorken, Baier, 0712.2456} is known.

The plan of the paper is as follows. In Section 2 we outline the conservative solutions in the Boltzmann equation. We then state and investigate our proposal for the regularity condition on the energy-momentum tensor for pure gravity in $AdS_{5}$ in Section 3. Finally, in the Discussion we point out the various novel and open issues that have been raised in the course of development of our results and proposal here; particularly those which we may hope to understand in the near future. The proof of existence of the conservative solutions in the Boltzmann equation is slightly technical and elaborate; we present this proof in detail in the Appendix in a self-contained manner.

\section{THE CONSERVATIVE SOLUTIONS OF THE BOLTZMANN EQUATION}
The study of equilibrium and transport properties of dilute gases through the dynamics of one-particle phase space distribution functions was pioneered by Maxwell \cite{Maxwell} and further developed by Boltzmann \cite{Boltzmann} in the 19th century. The Boltzmann equation provides a successful description of nonequilibrium phenomena in rarefied monoatomic gases. It is an equation for the evolution of the one-particle phase space distribution function. It can successfully describe nonequilibrium phenomena in rarefied gases, even at length scales between the microscopic molecular length scale and the mean-free path, and time scales between the time it takes to complete binary molecular collisions \cite{footnote4} and the average time between intermolecular collisions.

The Boltzmann equation is neither microscopic nor phenomenological, but a result of averaging the dynamics over microscopic length scales and time scales. Unlike phenomenological equations, it has no undetermined parameters and is completely fixed once the intermolecular force law is known. The structural details of the molecules are however ignored and effectively they are taken to be pointlike particles. The hydrodynamic equations with all the transport coefficients can be determined from the Boltzmann equation.

We start with a brief description of the conservative solutions of the Boltzmann equation for a system of pointlike classical nonrelativistic particles interacting via a central force. As mentioned in the Introduction, the proof of existence and uniqueness of such solutions is detailed in the Appendix in a self-contained manner. This is followed by a discussion on how to generalize our construction of conservative solutions to the semiclassical and relativistic versions of the Boltzmann equation. Finally we show how our results apply to multicomponent systems relevant for relativistic gauge theories. These generalizations are straightforward and the discussion on the nonrelativistic Boltzmann equation will be convenient for a first understanding of the conservative solutions.

\subsection{The conservative solutions in brief}
A generic solution of Boltzmann equation (\ref{be}) is characterized by infinite number of local variables. In general, these could be chosen to be the infinite local velocity moments ($f^{(n)}(\mathbf{x})$'s) of the one-particle phase space distribution $f(x,\xi)$, given by
\begin{equation}
f^{(n)}_{i_{1}i_{2}....i_{n}}(\mathbf{x},t)= \int d\xi\ \ c_{i_{1}}c_{i_{2}}.....c_{i_{n}} \ f(\mathbf{x},\xi) \quad .
\end{equation}
where $c_{i} = \xi_{i} - u_{i}(\mathbf{x}, t)$ with $u_{i}(\mathbf{x}, t)$ being the local average velocity.

However the first ten velocity moments suffice to parametrize the energy-momentum tensor. The conservative solutions, which are determined by the energy-momentum tensor alone, are thus a very special class of solutions obtained when the initial value data satisfy certain constraints.

Another special class of solutions to the Boltzmann equation is actually well known in the literature. These are the \emph{normal} solutions, where the local hydrodynamic variables given by the first five velocity moments of $f$ suffice to describe the solution even when it is far from equilibrium. Our conservative solutions are a generalization of these normal solutions. We review the normal solutions below before describing the conservative solutions.

\subsubsection{The hydrodynamic equations and normal solutions}
It is well-known that the first five velocity moments of the Boltzmann equation (\ref{be}), obtained by multiplying with $(1, \xi_{i}, \xi^{2})$ and integrating over $\xi$, give the hydrodynamic equations as 
\begin{eqnarray}\label{5mom}
\frac{\partial \rho}{\partial t} + \frac{\partial}{\partial x_{r}}(\rho u_{r}) &=& 0 \quad , \nonumber\\
\frac{\partial u_{i}}{\partial t} + u_{r} \frac{\partial u_{i}}{\partial x_{r}} + \frac{1}{\rho}\frac{\partial (p\delta_{ir}+p_{ir})}{\partial x_{r}} &=& 0 \quad , \\
\frac{\partial p}{\partial t} + \frac{\partial}{\partial x_{r}}(u_{r}p) +\frac{2}{3}(p\delta_{ir}+p_{ir})\frac{\partial u_{i}}{\partial x_{r}} +\frac{1}{3}\frac{\partial S_{r}}{\partial x_{r}} &=& 0 \quad , \nonumber
\end{eqnarray}
where the hydrodynamic variables $(\rho, u_{i}, p)$ are respectively the local density, components of local average molecular velocity and the local pressure of the gas defined in terms of the average root mean square kinetic energy. In terms of the velocity moments
\begin{eqnarray}\label{hydrovar}
\rho (\mathbf{x}, t) &=& \int f d\xi \quad , \nonumber\\
u_{i} (\mathbf{x}, t) &=& \frac{1}{\rho} \int \xi_{i} f d\xi \quad , \\
p (\mathbf{x}, t) &=& \frac{1}{3}\int \xi^{2} f d\xi \quad . \nonumber
\end{eqnarray}
The local temperature is defined through the local equation of state, $(RT = p/\rho)$ \cite{footnote}. The shear-stress tensor $p_{ij}$ and the heat flow vector $S_i$ (defined through $S_i = S_{ijk}\delta_{jk}$) are related to the velocity moments by
\begin{eqnarray}
p_{ij} &=& \int (c_{i}c_{j} - RT\delta_{ij}) f d\xi \quad , \nonumber\\
S_{ijk} &=& \int c_{i}c_{j}c_{k} f d\xi \quad ,
\end{eqnarray}
where $c_i = \xi_i - u_i$. It can be easily seen from the definition that $p_{ij} \delta_{ij} = 0$.

The collision term $J(f,f)$ (as defined in (\ref{J})) does not contribute when deriving the hydrodynamic equations (\ref{5mom}) from the Boltzmann equation. The first five velocity moments of $J(f,f)$ are zero owing to particle number, momentum and energy conservation as proven in the Appendix.

It must be emphasized that, in the hydrodynamic equations (\ref{5mom}), the shear-stress tensor $p_{ij}$ and the heat flow vector $S_{i}$ are functionally independent of the hydrodynamic variables. However there exist unique algebraic solutions to these and all the higher moments $f^{(n)}_{i_{1...}i_{n}}(\mathbf{x},t)$, which are functionals of the hydrodynamic variables. These functional forms contain only spatial derivatives of the hydrodynamic variables and can be systematically expanded in the so-called derivative expansion discussed below. This leads to the construction of the normal or purely hydrodynamic solutions of the Boltzmann equation, which we discuss below. For a generic solution of the Boltzmann equation, the higher moments of $f$ will have explicit time-dependent parts which are functionally independent of the hydrodynamic variables.

The $normal$ solutions of the Boltzmann equation \cite{Enskog,Burnett,Chapman} have been extensively discussed in \cite{book}. These solutions can be determined in terms of the five hydrodynamic variables $(\rho, u_i, p)$ \emph{alone}. They describe situations far away from equilibrium, such that observables which vanish at equilibrium do not vanish anymore but are functionally determined in terms of the hydrodynamic variables and their spatial derivatives. The existence of such solutions follows from the existence of unique algebraic solutions (as functionals of the hydrodynamic variables) to the equations of motion of the higher moments.
The functional forms of the shear-stress tensor and the heat flow vector, for instance, are given by
\begin{eqnarray}\label{funcpS}
p_{ij} &=& \eta \sigma_{ij} + \beta_{1}\frac{\eta^2}{p}(\partial \cdot u)\sigma_{ij}
+ \beta_{2}\frac{\eta^2}{p}\left(\frac{D}{Dt}\sigma_{ij} - 2\left(\sigma_{ik}\sigma_{kj} - \frac{1}{3} \delta_{ij}\sigma_{lm}\sigma_{lm}\right)\right)\nonumber\\
&&+ \beta_{3}\frac{\eta^2}{\rho T}\left(\partial_{i} \partial_{j} T - \frac{1}{3}\delta_{ij} \square T\right) + \beta_{4}\frac{\eta^2}{p\rho T}\left(\partial_{i} p \partial_{j} T + \partial_{j} p \partial_{i} T - \frac{2}{3} \delta_{ij} \partial_{l} p \partial_{l} T\right)\nonumber\\
&&+\beta_{5}\frac{\eta^2}{p\rho T}\left(\partial_{i} T \partial_{j} T - \frac{1}{3} \delta_{ij} \partial_{l} T \partial_{l} T\right) + ... \quad ,\\
S_{i} &=& \chi \partial_{i} T + ... \quad ,\nonumber
\end{eqnarray}
with the convective derivative $D/Dt = \partial/\partial t + u_{i}\partial_{i}$, and
\begin{eqnarray*}
\sigma_{ij} &=& \partial_{i}u_{j} + \partial_{j}u_{i} - \frac{2}{3}\delta_{ij}\partial \cdot u \quad , \nonumber\\
\eta &=& \frac{p}{B^{(2)}} + ... \quad ,   \qquad  \chi = \frac{15 R}{2}\eta + ...\quad ,
\end{eqnarray*}
where $\eta$ and $\xi$, appearing as in the Navier-Stokes equation and the Fourier's law of heat conduction, are the shear viscosity and heat conductivity respectively. $B^{(2)}$ is a specific function of the local thermodynamic variables determined by the collision kernel of the Boltzmann equation. The $\beta_{i}'$s are pure numbers that can be determined from the Boltzmann equation. The time derivative in $D/Dt$ can be converted to spatial derivatives using the hydrodynamic equations of motion; in fact, up to the orders shown above, we can assume that the Euler equation is valid and that the heat conduction is adiabatic.

The functional forms can be expanded systematically in the derivative expansion, which counts the number of spatial derivatives present in the expansion. The expansion parameter is the ratio of the typical length scale of variation of the hydrodynamic variables with the mean-free path. This is true for all the higher moments of $f$. The functional forms of $p_{ij}$ and $S_{i}$ (\ref{funcpS}), when substituted into the hydrodynamic equations (\ref{5mom}), give us systematic corrections to the Navier-Stokes equation and Fourier heat conduction respectively which can be expanded in the derivative expansion scheme.

The hydrodynamic equations are now the only dynamical equations. The higher moments are given algebraically in terms of the hydrodynamic variables and their spatial derivatives. The phase space distribution function $f$ is completely determined by the hydrodynamic variables through its velocity moments. The hydrodynamic equations thus form a closed system of equations and any solution of this system can be lifted to a unique solution of the full Boltzmann equation.

Stewart has shown \cite{reltboltzmann} that such normal solutions exist even for the relativistic and semiclassical Boltzmann equations.

\subsubsection{The conservative solutions}
We are able to prove that a more general class of special solutions to the Boltzmann equation - which we call \emph{conservative solutions} - exist. Here we outline these solutions, leaving the details of the proof to the Appendix. These solutions can be completely determined in terms of the energy-momentum tensor, analogous to the $normal$ solutions being completely determined in terms of the hydrodynamic variables. The energy-momentum tensor (as shown later) can be parametrized by the first ten moments of $f$ :
\begin{itemize}
\item
i) the five hydrodynamic variables ($\rho, u_{i}, p$), and
\item
ii) the five components of the shear-stress tensor $p_{ij}$ in a comoving locally inertial frame.
\end{itemize}
Importantly, for a generic conservative solution the shear-stress tensor is an \emph{independent variable} unlike the case of normal solutions, where it is a functional of the hydrodynamic variables.

These ten independent variables satisfy the following equations of motion
\begin{eqnarray}\label{9mom}
\frac{\partial \rho}{\partial t} + \frac{\partial}{\partial x_{r}}(\rho u_{r}) &=& 0 \quad ,\nonumber\\
\frac{\partial u_{i}}{\partial t} + u_{r} \frac{\partial u_{i}}{\partial x_{r}} + \frac{1}{\rho}\frac{\partial (p\delta_{ir}+p_{ir})}{\partial x_{r}} &=& 0 \quad ,\nonumber\\
\frac{\partial p}{\partial t} + \frac{\partial}{\partial x_{r}}(u_{r}p) +\frac{2}{3}(p\delta_{ir}+p_{ir})\frac{\partial u_{i}}{\partial x_{r}} +\frac{1}{3}\frac{\partial S_{r}}{\partial x_{r}} &=& 0 \quad ,\\
\frac{\partial p_{ij}}{\partial t} +\frac{\partial}{\partial x_{r}}(u_{r}p_{ij}) +\frac{\partial S_{ijr}}{\partial x_{r}} -\frac{1}{3}\delta_{ij}\frac{\partial S_{r}}{\partial x_{r}} \nonumber\\
+\frac{\partial u_{j}}{\partial x_{r}} p_{ir}+ \frac{\partial u_{i}}{\partial x_{r}} p_{jr} -\frac{2}{3}\delta_{ij}p_{rs}\frac{\partial u_{r}}{\partial x_{s}} \nonumber\\
+p(\frac{\partial u_{i}}{\partial x_{j}}+\frac{\partial u_{j}}{\partial x_{i}}-\frac{2}{3}\delta_{ij}\frac{\partial u_{r}}{\partial x_{r}}) &=& \sum_{p,q = 0; p \geq q; (p,q)\neq(2,0)}^{\infty} B^{(2,p,q)}_{ij\nu\rho}(\rho, T)f^{(p)}_{\nu}f^{(q)}_{\rho} \nonumber\\
&& +B^{(2)}(\rho, T)p_{ij} \quad .\nonumber
\end{eqnarray}
where $B^{(2,p,q)}_{ij\nu\rho}$ are determined by the collision kernel in the Boltzmann equation. $\nu$ and $\rho$ indicate abstractly all the $p$ and $q$ indices of the moments $f^{(p)}$ and $f^{(q)}$, respectively.

The above equations are now a closed system of equations, just like the hydrodynamic equations were in case of the normal solutions. All the higher moments appearing in the above equations are given as functionals of the hydrodynamic variables $and$ the stress tensor. These functional forms are unique and special algebraic solutions of the higher moments of $f$. For instance, the heat flow vector can be determined from
\begin{equation}
S_{i} = \frac{15pR}{2B^{(2)}}\frac{\partial T}{\partial x_{i}} + \frac{3}{2B^{(2)}}\left(2RT\frac{\partial p_{ir}}{\partial x_{r}}+ 7R p_{ir}\frac{\partial T}{\partial x_{r}} - \frac{2p_{ir}}{\rho}\frac{\partial p}{\partial x_{r}}\right)+.... \quad .
\end{equation}
The functional forms of all the higher moments, as for the heat flow vector above, can be expanded systematically in two expansion parameters $\epsilon$ and $\delta$. The parameter $\epsilon$ is the old derivative expansion parameter -- the ratio of the typical length scale of spatial variation to the mean-free path. The new parameter $\delta$ is an amplitude expansion parameter, defined as the ratio of the typical amplitude of the nonhydrodynamic shear-stress tensor with the hydrostatic pressure in the final equilibrium.

The closed system of ten equations (\ref{9mom}) are thus the only dynamical equations and any solution of this system can be lifted to a full solution of the Boltzmann equation through the unique functional forms of the higher moments.

The normal solutions,being independent of the stress tensor, are clearly a special class of conservative solutions. There is another interesting class of conservative solutions which are homogenous or invariant under spatial translations. The phase space distribution function $f$ is a function of $\mathbf{v}$ only for these homogenous solutions and the hydrodynamic variables are constants both over space and time [this can be easily seen from (\ref{9mom})]. The shear-stress tensor and consequently all the higher moments are functions of time alone. Such solutions have dynamics in velocity space only and describe relaxation processes.

In a generic solution of the Boltzmann equation, the dynamics at short time scales is more like the homogenous class, where the initial one-particle distribution relaxes to a local equilibrium given by a local Maxwellian distribution parametrized by the local values of the hydrodynamic variables. At long time scales the dynamics is more like the normal solutions, where the system goes to global equilibrium hydrodynamically. Thus conservative solutions, despite being mathematically special, capture both relaxation and hydrodynamics which constitute generic nonequilibrium processes in a phenomenological manner. In other words, the dynamics of the energy-momentum tensor alone given by (\ref{9mom}) captures both relaxation and hydrodynamics in a systematic fashion.

\subsection{Relativistic and semiclassical corrections to conservative solutions}
The proof for existence of conservative solutions in the nonrelativistic classical Boltzmann equation can be readily generalized to its semiclassical and relativistic versions. This is because all the properties of the collision term $J$ required for the proof of the existence of conservative solutions carry over to the semiclassical and relativistic versions as well.

Let us consider the semiclassical version of the collision term which takes into account quantum statistics. This was first obtained by Uehling and Uhlenbeck \cite{Uhlenbeck} to be
\begin{eqnarray}
J(f,g) &=& \int \mathcal{J}(\xi, \xi^{*}) B(\theta, V)d\xi^{*}d\epsilon d\theta\ \quad , \nonumber\\
\mathcal{J}(\xi, \xi^{*}) &=& \left[f(\mathbf{x,\xi^{'}})g(\mathbf{x,\xi^{*'}}) \mathcal{F}(\xi) \mathcal{G} (\xi^{*}) -f(\mathbf{x,\xi})g(\mathbf{x,\xi^{*}}) \mathcal{F}(\xi^{'}) \mathcal{G} (\xi^{*'})\right] \quad ,\nonumber\\
\mathcal{F}(\xi) &=&  \left(1 \pm \frac{h^{3}f(\xi^{'})}{(2s+1)}\right), \quad \mathcal{G}(\xi) = \left(1 \pm \frac{h^{3}g(\xi)}{(2s+1)}\right) \quad ,
\end{eqnarray}
where the $+$ sign applies for bosons, the $-$ sign for fermions and $s$ is the spin of the particles comprising the system. The final velocities $\xi^{'}$ and $\xi^{*'}$ are determined by the velocties $\xi$ and $\xi^{*}$ before the binary molecular collision according to the intermolecular force law. Importantly, now $J(f,f)$ vanishes if and only if $f$ is the Bose-Einstein or the Fermi-Dirac distribution in velocity space for bosons and fermions respectively, instead of being the Maxwellian distribution \cite{footnote6}.

The proof for the existence of conservative solutions in the nonrelativistic classical case does not require any explicit form of the collision kernel $J$. Only certain key properties suffice, as will be evident from the proof. We can pursue the same strategy with the semiclassical corrections as well \cite{footnote7}.

One has to employ the Sonine polynomials, which are generalizations of Hermite polynomials, to find solutions of the required algebraic solutions of the higher moments as in \cite{Rupak}. The main objection could be that for the proof of existence of solutions, we use a theorem due to Hilbert which is explicitly stated for the nonrelativistic classical $J$. However the details are exactly the same as that for constructing the normal solutions. It has been seen that normal solutions can indeed be constructed in the semiclassical case \cite{reltboltzmann}, so there ought to be no obstruction to the construction of conservative solutions also. Indeed, our proof shows that we can construct the conservative solutions given that the normal solutions exist.

The generalization in the relativistic case again holds on similar grounds as above. It is more convenient to use a covariant description now. Normal solutions of the semiclassical relativistic Boltzmann equation have also been constructed \cite{reltboltzmann}. So there should be no obstruction in constructing conservative solutions as well.

In fact the same arguments could be used to state that any solution of the relativistic semiclassical Boltzmann equation at sufficiently late times can be approximated by an appropriate conservative solution, since the maximum speed of propagation of linearized modes increases monotonically \cite{mullerrev} as more and more higher moments are included.

\subsection{Multicomponent systems}

So far we have pretended as if our system is composed of only one component or particle. However gauge theories have many species of particles and internal degrees of freedom, hence we need to understand how to extend our results to multicomponent systems.

Let us consider the example of $\mathcal{N}=4$ super Yang-Mills theory. In the weakly coupled description we need to deal with all the adjoint fermions and scalars along with the gauge bosons; all these particles form a SUSY multiplet. We note that in the universal sector all charge densities or currents corresponding to local (gauge) and global [$SO(6)_{R}$] charges are absent. Similarly we should not have any multipole moments of local or global charge distributions, because in the gravity side we have pure gravity only. Therefore, most naturally we should have that all members of the $\mathcal{N} = 4$ SUSY multiplet, distinguished by their spin, global charge and color, should be present in equal density at all points in phase space. So we are justified in our analysis in dealing with a single phase space distribution $f$. The Boltzmann equation we have considered above is obtained after summing over interactions in all possible spin, charge and color channels.

The situation should be similar in any other conformal gauge theory. We can still treat the spin, color and charge as internal degrees of freedom owing to mass degeneracy even though the particles do not form a SUSY multiplet. In the absence of any chemical potential, there should be equipartition at all points in phase space over these internal degrees of freedom. This should be the most natural weak coupling extrapolation of the situation in the universal sector, dual to pure gravity, where gravity is blind to all the internal degrees of freedom of the particles.

\section{APPLICATIONS TO PURE GRAVITY IN ADS}
We will now argue that conservative solutions should exist even in the exact microscopic theory. In the exact microscopic theory, we do not make any approximation over the microscopic degrees of freedom and their dynamics unlike the Boltzmann equation, though an appropriate averaging over the environmental degrees of freedom is required to get the final equilibrium configuration.

To begin with, consider the BBGKY heirarchy of equations \cite{BBGKY} which describes a hierarchy of coupled  semiclassical nonrelativistic equations for the evolution of multiparticle phase space distributions. This hierarchy is useful for developing kinetic theory of liquids. If the hierarchy is not truncated, then it is equivalent to the exact microscopic description. It has been shown that normal or purely hydrodynamic solutions to the untruncated hierarchy exist. These solutions lead to the determination of viscosity of liquids which behave correctly as a function of density and temperature \cite{Normal}. It is therefore likely that the conservative solutions also exist for this system which means they are likely to exist for the microscopic nonequilibrium theory of nonrelativistic classical systems constituted by pointlike particles.

Experiments at the Relativistic Heavy Ion Collider (RHIC) suggest that the evolution of quark-gluon plasma (QGP) can be well approximated by hydrodynamic equations, very soon after its formation from the fireball \cite{Heinz}. Given that the perturbative nonequilibrium dynamics of hot QCD for temperatures greater than the microscopic scale $\Lambda$ is equivalent to a relativistic semiclassical Boltzmann equation \cite{Arnold}, we know perturbatively normal or purely hydrodynamic solutions exist for these microscopic theories. In fact any generic solution of the relativistic semiclassical Boltzmann equation can be approximated by an appropriate normal solution at a sufficiently late time. The quick approach to almost purely hydrodynamic behavior for the strongly coupled QGP in RHIC suggests that  even nonperturbatively a normal solution should exist which could approximate the late-time behavior for any generic nonequilibrium state. It is also true that not all transport coefficients of generic conformal  higher derivative hydrodynamics can be defined through linear response theory. The plausible route of defining these higher order transport coefficients could be through the construction of normal solutions in nonequilibrium quantum field theories. Extremely fast relaxation dynamics in quark-gluon plasma similarly suggest that conservative solutions should capture generic nonequilibrium behavior. This is because in such systems the approach to the conservative regime, where the dynamics is given in terms of the energy-momentum tensor alone, should be faster than in weakly coupled systems, where even the corrections to Navier-Stokes hydrodynamics are hard to determine experimentally.

If we accept that conservative solutions exist in the exact microscopic theory, it is only natural to identify the conservative solutions with the universal sector at large $N$ and strong coupling in gauge theories with gravity duals. Such an identification explains the dynamics in the universal sector being determined by the energy-momentum tensor alone. We emphasize, however, that the conservative solutions become universal only at strong coupling and large $N$.

An appropriate AdS/CFT argument can also be provided for the existence of conservative solutions for finite $N$ and coupling. In such cases, we need to consider higher derivative corrections to Einstein's equation and the full $ten-dimensional$ equations of motion. There is no guarantee of a consistent truncation to pure gravity anymore. However, we can use holographic renormalization with Kaluza-Klein reduction to five dimensions \cite{kkhol} to argue that we can readily extend the solutions in the universal sector, perturbatively in the string tension ($\approx 1/\sqrt{\lambda}$ in appropriate units) and string coupling (whose $N$ dependence is $1/N$). This can be done by turning off the normalizable mode of the dilaton while keeping its non-normalizable mode constant, turning off the normalizable and non-normalizable modes of all other fields while keeping the boundary metric flat and perturbatively correcting the energy-momentum tensor to appropriate orders of the string tension and string coupling, so that the gravity solution still has a future horizon regular up to desired orders in the perturbation expansion. These solutions, again by construction, are determined by energy-momentum tensor alone. Our claim that the conservative solutions exist in the exact microscopic theory at any value of coupling and $N$ is therefore validated.

The identification of conservative solutions with the universal sector at strong coupling and large $N$ for conformal gauge theories with gravity duals allows us to  create a framework for solutions of pure gravity in AdS with regular future horizons. We first study the parametrization of the boundary stress tensor which will allow us to make the connection with nonequilibrium physics. Then we will proceed to give a framework for regular solutions, with the only assumption being the identification of conservative solutions with the universal sector at strong coupling and large $N$. Finally we will make some connections with known results.

\subsection{The energy-momentum tensor and nonequilibrium physics}

A general parametrization of the energy-momentum tensor allows us to connect gravity to the nonequilibrium physics of conformal gauge theories. This parametrization has been first applied in the AdS/CFT context in\cite{Baier}. The energy-momentum tensor is first written as
\begin{equation}\label{split1}
t_{\mu\nu} = t_{(0)\mu\nu} + \pi_{\mu\nu} \quad ,
\end{equation}
where $t_{(0)\mu\nu}$ is the part of the energy-momentum tensor in local equilibrium. It can be parametrized in conformal theories by the hydrodynamic variables, the timelike velocity ($u^{\mu}$) and the temperature ($T$), as

\begin{equation}\label{split2}
t_{(0)\mu\nu} = (\pi T)^{4} (4u_{\mu}u_{\nu}+\eta_{\mu\nu}) \quad ,
\end{equation}
and $\pi_{\mu\nu}$ is the nonequilibrium part of the energy-momentum tensor.

If we define the four velocity $u^{\mu}$ to be the local velocity of energy transport and the temperature $T$ such that $3(\pi T)^{4} = u^{\mu}u^{\nu}t_{\mu\nu}$ is the local energy density, then in the local frame defined through $u^{\mu}$, the energy-momentum tensor must receive nonequilibrium contributions in the purely spatial block orthogonal to the four velocity. This means
\begin{equation}\label{split3}
u^{\mu}\pi_{\mu\nu} = 0 \quad .
\end{equation}

The constraints in Einstein's equations impose the tracelessness and conservation condition on the energy-momentum tensor so that
\begin{eqnarray}\label{split4}
\partial^{\mu}t_{\mu\nu}=0 \quad \Rightarrow \quad \partial^{\mu}\left((\pi T)^{4} (4u_{\mu}u_{\nu}+\eta_{\mu\nu})\right) &=& - \partial^{\mu}\pi_{\mu\nu} \quad ,\nonumber\\
Tr(t) = 0  \Rightarrow Tr(\pi) &=& 0 \quad .
\end{eqnarray}
In the second equation above, the implication for the tracelessness for $\pi_{\mu\nu}$ comes from the
fact that the equilibrium energy-momentum tensor as given by (\ref{split2}) is by itself traceless.

In the dual theory these conditions are satisfied automatically owing to the full $SO(4,2)$ conformal invariance. Note that the first of the equations above is just the forced Euler equation and can be thought of as the equation of motion for the hydrodynamic variables. 

We can reinterpret a class of known solutions of pure gravity in AdS as the duals of the normal solutions in the exact microscopic theory at strong coupling and large $N$.  These solutions are the "tubewise black-brane solutions" \cite{0712.2456} which, in any radial tube ending in a patch at the boundary, are approximately boosted black brane solutions corresponding to local equilibrium and can be parametrized by the hydrodynamic variables corresponding to the patch at the boundary. These solutions can be constructed perturbatively in the derivative expansion. The expansion parameter, being the ratio of length and time scale of variation of the local hydrodynamic parameters and the mean-free path in final equilibrium, simply counts the number of boundary derivatives. We can identify these solutions as duals of normal solutions because the nonequilibrium part of the energy-momentum tensor $\pi_{\mu\nu}$ can be parametrized by the hydrodynamic variables and their derivatives alone.

The complete parametrization of the purely hydrodynamic $\pi_{\mu\nu}$ in any conformal theory is known up to second order in the derivative expansion.In this parametrization, aside from the shear viscosity four higher order transport coefficients appear \cite{Baier, 0712.2456}, which can be fixed by requiring the regularity of the future horizon giving us the tubewise black-brane solutions \cite{0712.2456}.

Let us define the projection tensor $P_{\mu\nu}$ which projects on the spatial slice locally orthogonal to the velocity field, so that
\begin{eqnarray}
P_{\mu\nu} = u_{\mu}u_{\nu} + \eta_{\mu\nu} \quad .\nonumber
\end{eqnarray}
The hydrodynamic shear strain rate $\sigma_{\mu\nu}$ is defined as
\begin{equation}
\sigma^{\mu\nu} = \frac{1}{2}P^{\mu\alpha}P^{\nu\beta}\left(\partial_{\alpha}u_{\beta}+\partial_{\beta}u_{\alpha}\right) -\frac{1}{3} P^{\mu\nu}(\partial \cdot u) \quad .
\end{equation}
We also introduce the hydrodynamic vorticity tensor,
\begin{equation}
\omega^{\mu\nu} = \frac{1}{2}P^{\mu\alpha}P^{\nu\beta}(\partial_{\alpha}u_{\beta}-\partial_{\beta}u_{\alpha}) \quad .
\end{equation}

The purely hydrodynamic $\pi_{\mu\nu}$ up to second order in the derivative expansion, for the tubewise black-brane solutions, with all nonvanishing transport coefficients fixed by regularity is
\begin{eqnarray}\label{soh}
\pi^{\mu \nu} &=& - 2(\pi T)^3 \sigma^{\mu \nu} \nonumber\\
&&+ (2 - \ln 2) (\pi T)^2 \left[(u \cdot \partial) \sigma^{\mu \nu} + \frac{1}{3}\sigma^{\mu \nu}(\partial \cdot u) - (u^{\nu} \sigma^{\mu \beta}  + u^{\mu} \sigma^{\nu \beta}) (u \cdot \partial)u_{\beta}\right] \nonumber\\
&&+ 2 (\pi T)^2 \left(\sigma^{\alpha\mu}\sigma_{\alpha}^{\phantom{\alpha}\nu} - \frac{1}{3}P^{\mu \nu} \sigma_{\alpha \beta}\sigma^{\alpha \beta}\right) \nonumber\\
&&+(\ln 2)(\pi T)^2  (\sigma^{\alpha\mu}\omega_{\alpha}^{\phantom{\alpha}\nu}+\sigma^{\alpha\mu}\omega_{\alpha}^{\phantom{\alpha}\nu}) + O(\partial^3 u) \quad .
\end{eqnarray}

Having identified the normal solutions in the universal sector with a class of solutions which could in principle be constructed up to any order in the derivative expansion, we will now naturally extend this observation to a framework which captures all regular solutions in certain expansion parameters.

\subsection{The complete framework}

In the hydrodynamic case we had four hydrodynamic variables, so the conservation of the energy-momentum tensor alone is sufficient to determine the evolution in the boundary. However in the generic case we need an independent equation of motion for $\pi_{\mu\nu}$.

The regularity condition must be an equation for the evolution of $\pi_{\mu\nu}$ similar to the last equation of (\ref{9mom}), This is because, as per our argument, the conservative solutions should be identified with the universal sector at large $N$ and strong coupling. However Eq. (\ref{9mom}) came from an underlying Boltzmann equation. At strong coupling, we have no kinetic equation to guide us because a valid quasiparticle description at strong coupling is not known even for $\mathcal{N}=4$ supersymmetric Yang-Mills theory. Moreover an entropy current cannot be probably constructed beyond the class of purely hydrodynamic solutions, hence we cannot use any formalism like the Israel-Stewart-Muller formalism \cite{ISM} to guess an equation for $\pi_{\mu\nu}$. This is because we should not expect a monotonic approach to equilibrium, as in the case of the Boltzmann equation, when we go to the exact microscopic description \cite{footnote14}.

The safest strategy therefore, will be to use only the following basic inputs without resorting to guesswork.
\begin{itemize}
\item
The first input is that the equation for $\pi_{\mu\nu}$ has to be conformally covariant because the dual gauge theory is conformal.
\item
The second input is that the solutions in the purely hydrodynamic sector are known exactly up to second order in the derivative expansion and, being identified with the normal solution, should be special cases of our complete framework. The equation for $\pi_{\mu\nu}$ must therefore have (\ref{soh}), the purely hydrodynamic energy-momentum tensor known up to second order in the derivative expansion, as a solution up to those orders.
\end{itemize}

With only these inputs, we will be able to propose the equation for $\pi_{\mu\nu}$ only up to certain orders of expansion in both the hydrodynamic and nonhydrodynamic expansion parameters about the equilibrium state. However we should consider the most general equation for $\pi_{\mu\nu}$ which satisfies the above criteria. The expansion parameters are again the derivative expansion parameter (as in the purely hydrodynamic sector, but with the spatio-temporal variation of $\pi_{\mu\nu}$ taken into account additionally) and the amplitude expansion parameter, which is the ratio of a typical value of $\pi_{\mu\nu}$ divided by the pressure in final equilibrium.

Our proposal then amounts to the following equation of motion for $\pi_{\mu\nu}$, whose solutions should give all the regular solutions of pure gravity in $AdS_{5}$ :
\begin{eqnarray}\label{proposal1}
(1-\lambda_{3})\left[(u\cdot\partial)\pi^{\mu\nu} + \frac{4}{3}\pi^{\mu\nu}(\partial\cdot u) - \left(\pi^{\mu \beta} u^{\nu} + \pi^{\nu \beta} u^{\mu}\right) (u \cdot \partial)u_{\beta}\right] \nonumber\\
= -\frac{2\pi T}{(2-\ln 2)}\Bigg[\pi^{\mu\nu}+2(\pi T)^{3}\sigma^{\mu\nu} \nonumber\\
-\lambda_{3}(2-\ln2)(\pi T)^{2}\left((u\cdot\partial)\sigma^{\mu\nu} + \frac{1}{3}\sigma^{\mu\nu}(\partial\cdot u) - \left(u^{\nu} \sigma^{\mu \beta} + u^{\mu} \sigma^{\nu \beta}\right) (u\cdot\partial)u_{\beta}\right) \nonumber\\
- \lambda_{4}(\ln 2)(\pi T)^{2}(\sigma^{\alpha\mu}\omega_{\alpha}^{\phantom{\alpha}\nu}+\sigma^{\alpha\mu}\omega_{\alpha}^{\phantom{\alpha}\nu}) \nonumber\\
-2\lambda_{1}(\pi T)^{2}\left(\sigma^{\alpha \mu}\sigma^{\nu}_{\phantom{\nu}\alpha}- \frac{1}{3}P^{\mu\nu}\sigma^{\alpha\beta}\sigma_{\alpha\beta}\right)\Bigg] \nonumber\\
- (1-\lambda_{4})\frac{ln 2}{(2- \ln 2)}(\pi^{\mu}_{\phantom{\nu}\alpha}\omega^{\alpha\nu} + \pi^{\nu}_{\phantom{\nu}\alpha}\omega^{\alpha\mu})  \nonumber\\ -\frac{2 \lambda_{2}}{(2- \ln 2)}\left[\frac{1}{2}(\pi^{\mu\alpha}\sigma^{\nu}_{\phantom{\mu}\alpha}+\pi^{\nu\alpha}\sigma^{\mu}_{\phantom{\mu}\alpha})- \frac{1}{3}P^{\mu\nu}\pi^{\alpha\beta}\sigma_{\alpha\beta}\right] \nonumber\\
+\frac{1 -\lambda_{1}-\lambda_{2}}{(2- \ln 2)(\pi T)^{3}}\left(\pi^{\mu\alpha}\pi^{\nu}_{\phantom{\nu}\alpha}-\frac{1}{3}P^{\mu\nu}\pi^{\alpha\beta}\pi_{\alpha\beta}\right) \nonumber\\ + O\left(\pi^{3}, \pi\partial\pi, \partial^{2}\pi, \pi^{2}\partial u, \pi\partial^{2}u, \partial^{2}\pi, \partial^{3}u, (\partial u)(\partial^{2}u), (\partial u)^{3}\right) \quad ,
\end{eqnarray}
where the $O(\pi^{3}, \pi\partial\pi, ...)$ term indicates the corrections which lie beyond our inputs. 

The corrections can only include terms of the structures displayed or those with more derivatives or containing more powers of $\pi_{\mu\nu}$ or both. We cannot say much about these corrections because for purely hydrodynamic solutions, they contribute to the energy-momentum tensor at the third derivative order only and the general structure of the hydrodynamic energy-momentum tensor at third order in derivatives is not known. The four $\lambda_{i}$'s ($i=1,2,3,4$) are pure numbers. Though we have not been able to specify their values, they are not free parameters. Once their
values are fixed by regularity of the future horizon for certain configurations, they should give the complete framework for the whole class of regular solutions.

As we have already mentioned, this equation of motion (\ref{proposal1}) for the shear-stress tensor $\pi_{\mu\nu}$ has to be supplemented by the conservation of energy-momentum tensor in the form given in (\ref{split4}) so that we have nine equations for the nine variables (including the hydrodynamic variables) parametrizing the general nonequilibrium energy-momentum tensor. The tracelessness of the energy-momentum tensor begets the tracelessness of $\pi_{\mu\nu}$ as in (\ref{split4}) and this, as we have mentioned before, has led to the requirement that our equation of motion for $\pi_{\mu\nu}$ should be Weyl covariant.

This equation is thus a phenomenological framework for the universal sector as a whole up to certain orders in perturbation about the final equilibrium state. This framework governs both hydrodynamic and nonhydrodynamic situations and goes much beyond linear perturbation theory. This is however, only valid within the universal sector. Beyond this sector we need many other inputs other than the boundary energy-momentum tensor to specify the boundary state or the solutions in gravity.

\subsection{Checks, comparisons and comments}

We will begin with a couple of comments. The first comment is that our Eq. (\ref{proposal1}) does not hold well at early times in the generic case. At early times the terms with time derivatives of various orders coming from the higher order corrections to our equation would become important. We will soon see the effect of such time-derivative terms in a simple example. We give an argument why such terms with time-derivatives must appear in the higher order corrections \cite{thank1}. Any data at early times in the bulk, which will result in smooth behavior in the future, should get reflected in terms of an infinite set of variables in the boundary. The only way we can represent this in terms of the energy-momentum tensor alone is to include its higher order time derivatives in the initial data, so the equation for evolution of the energy-momentum tensor should contain higher order time derivatives.

The second comment is that, in the particular case of boost-invariant flow, we have a better structural understanding of the hydrodynamic behavior at higher orders in the derivative expansion \cite{Kinoshita}. We can, in principle, use our procedure to give a framework for general boost-invariant flows at late times. However we will leave this for future work. Moreover, the basic logic of our proposal is to use the purely hydrodynamic behavior as an input and then extend this to the complete framework. So our proposal and its extension at higher orders, by construction, reproduce the hydrodynamic sound and shear branches of the quasinormal modes.

We now develop a straightforward strategy to check our proposal. We could look at simple nonhydrodynamic configurations first and construct the bulk solution perturbatively in the amplitude expansion parameter to determine some of the $\lambda$'s. Once these have been determined, we can construct bulk solutions corresponding to a combination of hydrodynamic and nonhydrodynamic behaviors perturbatively in both the amplitude and derivative expansion parameters and then check if the regularity fixes those $\lambda$'s to the same values.

The simplest nonhydrodynamic configurations are the analogs of homogenous conservative solutions of the Boltzmann equation we have mentioned before and which describe pure relaxation dynamics. Such configurations are homogenous in space, but time dependent and satisfy the conservation equation trivially. In such configurations the flow is at rest, so that $u^{\mu}=(1,0,0,0)$ and the temperature $T$ is also a spatiotemporal constant. The nonequilibrium part of the energy-momentum tensor satisfies the following conditions \newline (i) the time-time component $\pi_{00}$ and the time-space components $\pi_{0i}$ for $i=1,2,3$ vanish and \newline (ii) the space-space components $\pi_{ij}$ for $i,j =1,2,3$ are dependent only on time.

The above conditions on $\pi_{\mu\nu}$ result in the conservation equation being trivially satisfied. It follows from our proposal (\ref{proposal1}) that regularity in the bulk implies that $\pi_{ij}$ satisfy the following equation of motion :
\begin{equation}\label{nh}
(1-\lambda_{3})\frac{d\pi_{ij}}{dt} + \frac{2\pi T}{(2- ln 2)}\pi_{ij} -\frac{1 -\lambda_{1}-\lambda_{2}}{(2- ln 2)(\pi T)^{3}}\left(\pi_{ik}\pi_{kj}-\frac{1}{3}\delta_{ij}\pi_{lm}\pi_{lm}\right) = O(\frac{d^{2}\pi_{ij}}{dt^{2}}) \quad .
\end{equation}

If we look at the linearized solution, we have
\begin{equation}\label{linsolve}
\pi_{ij} = \mathcal{A}_{ij}exp(-\frac{t}{\tau_{\pi}}), \quad \tau_{\pi} = (1 - \lambda_{3})\frac{2- ln \ 2}{2\pi T } \quad ,
\end{equation}
where $\mathcal{A}_{ij}$ is a spatiotemporally constant matrix such that $\mathcal{A}_{ij}\delta_{ij} = 0$. This implies that we have a nonhydrodynamic mode such that when the wave vector $\mathbf{k}$ vanishes, the frequency $\omega$ becomes purely imaginary and equals $-i\tau_{\pi}^{-1}$, i.e $\omega = -i\tau_{\pi}^{-1}$ as $\mathbf{k}\rightarrow 0$. There is however, no such mode in the quasinormal spectrum of black branes \cite{Kovtun}. This makes us conclude that $\lambda_{3} =1$, so that at the linearized level the only solution of (\ref{nh}) is $\pi_{ij}=0$.

However, at the nonlinear level we still have nonhydrodynamic solutions given by
\begin{equation}\label{snh}
\frac{2\pi T}{(2- ln 2)}\pi_{ij} -\frac{1 -\lambda_{1}-\lambda_{2}}{(2- ln 2)(\pi T)^{3}}\left(\pi_{ik}\pi_{kj}-\frac{1}{3}\delta_{ij}\pi_{lm}\pi_{lm}\right) = O(\frac{d^{2}\pi_{ij}}{dt^{2}}) \quad .
\end{equation}
In fact, up to the orders explicitly given above, the equation is nondynamical and predicts that we should, at least perturbatively, have tensor hair on the black-brane solution in pure gravity in AdS. This gives us the simplest nontrivial test of our proposal and also a means of determining $\lambda_{1} +\lambda_{2}$.

In this connection, we also note that the possible second order in the time-derivative correction in (\ref{snh}) implies that we need not have a monotonic approach to equilibrium as we should have in the presence of an entropy current.

We end here with some comments on the issue of connecting our proposal with physics of quasinormal modes of the black brane. The linearized limit of the conservation equation, along with our proposed Eq. (\ref{proposal1}), supports at most three branches of linearized fluctuations. We have further argued that the third branch giving pure relaxation dynamics is not present. However, we know that the quasinormal modes have infinite branches of higher overtones other than the hydrodynamic sound and shear branches. This naive comparison is somewhat misplaced \cite{thank2} because, as we know, nonlinearities do affect linearized propagation in quantum field theories. Since our equations are actually equivalent to the nonequilibrium field theory equation of motion of the state in the gauge theory, we must take into account nonlinearities of our equation in the propagation of the energy-momentum tensor before making any comparison. We leave this for future work.

We would also like to mention here that it is only natural that the higher overtones are more like resonances and are built out of the dynamics of the nine degrees of freedom of the conformal energy-momentum tensor, as it would have been surprising if infinite branches in the spectrum in the universal sector would have been blind to the microscopic details of the theory like the matter content and couplings. Our framework suggests that these infinite branches could be obtained from the nonlinear dynamics of the energy-momentum tensor. However we should exhibit caution here because although these nonhydrodynamic higher overtones of quasinormal modes are indeed regular linear perturbations of the black brane, it is yet to be demonstrated that these can be developed into complete regular solutions of Einstein's equation nonlinearly.

\section{DISCUSSION}

We will mention some of the developments on which we would like to focus in the future. The first could be in the realm of early-time dynamics, especially in the understanding of decoherence. This should be convenient because we can understand a lot by just considering the higher order corrections to the homogenous nonhydrodynamic configurations which solve (\ref{snh}). We have already observed the possibility of an oscillatory approach to equilibrium here. To uncover the physics, we need to compare with homogenous conservative solutions in quantum kinetic theories which can capture the physics of decoherence. We can test whether the same dynamics of the energy-momentum tensor in conservative solutions of quantum kinetic theories which captures decoherence, also gives rise to horizon formation in the bulk.

A second important issue would be a better understanding of whether the hydrodynamic limit of nonequilibrium dynamics always leads to generation of an entropy current. The tubewise black-brane solutions, which by our logic should constitute the normal or purely hydrodynamic solutions at large $N$ and strong coupling, indeed demonstrate the existence of a family of entropy currents \cite{entropy}. However, unlike for the normal solutions of the Boltzmann equation, or in the Israel-Stewart-Muller formalism, these entropy currents are not of the form $su^{\mu}$, where $s$ could be interpreted as the nonequilibrium entropy density. We hope to get a better understanding of the physics of this entropy current by investigating the existence and form of the entropy currents in the normal solutions of untruncated BBGKY heirarchy which, as mentioned before, are solutions of exact microscopic dynamics.

Finally, we have given a framework for general universal nonequilibrium behavior in strongly coupled gauge theories with gravity duals. It would be interesting to see how much of this framework may apply to physics of quark-gluon plasma at the RHIC.

\begin{flushleft}
\textbf{Acknowledgments :}
\end{flushleft}
We would like to thank Ashoke Sen, Rajesh Gopakumar, Dileep Jatkar, Justin David, Laurence Yaffe, A. O. Starinets, Eric Verlinde, Kostas Skenderis and Shiraz Minwalla for various valuable discussions. AM would like to acknowledge IISc, IMSc, the hosts of the Fifth Aegean Winter School, LPTHE, ICTP, the University of Amsterdam, the Universite libre de Bruxelles, Imperial College, DAMTP, Harvard University, Caltech and TIFR for invitations for giving seminars on this work prior to publication. AM would also like to thank the people of India for generously supporting this work.

\section*{APPENDIX : PROOF OF EXISTENCE OF CONSERVATIVE SOLUTIONS}
We will now present the details of our proof for the existence and uniqueness of conservative solutions of the Boltzmann equation. In order to keep the proof reasonably self-contained, we give further details on the Boltzmann equation and how one can obtain the hydrodynamic equations seen earlier. We follow the notation of \cite{Grad, book} mostly for this part of the discussion. This will be followed by the proof in full detail.
\subsection*{A.1\ \ A short description of the Boltzmann equation}
The Boltzmann equation for the one-particle phase space distribution $f(\mathbf{x,\xi})$ for a gas of nonrelativistic monoatomic molecules of unit mass interacting through a central force is
\begin{equation}\label{be}
(\frac{\partial}{\partial t} +\xi \cdot \frac{\partial}{\partial \mathbf{x}}) f(\mathbf{x,\xi}) = J(f,f)(\mathbf{x,\xi}) \quad ,
\end{equation}
where
\begin{equation}\label{J}
J(f,g) = \int \left(f(\mathbf{x,\xi^{'}})g(\mathbf{x,\xi^{*'}})-f(\mathbf{x,\xi})g(\mathbf{x,\xi^{*}})\right)B(\theta, V)d\xi^{*}d\epsilon d\theta \quad ,
\end{equation}
is the collision integral. ($\mathbf{\xi, \xi^{*}}$) are the velocities of the molecules before a binary collision and ($\mathbf{\xi^{'}, \xi^{*'}}$) are their corresponding velocities after the collision. The angular coordinates ($\theta, \epsilon$) are the coordinates related to the collision, and $\mathbf{V} = \mathbf{\xi} -\mathbf{\xi}^{*}$ is the relative velocity with magnitude $V$. We assume that the collision takes place due to a central force acting between the molecules.

\begin{figure}[!h]
\begin{center}
\epsfig{file = 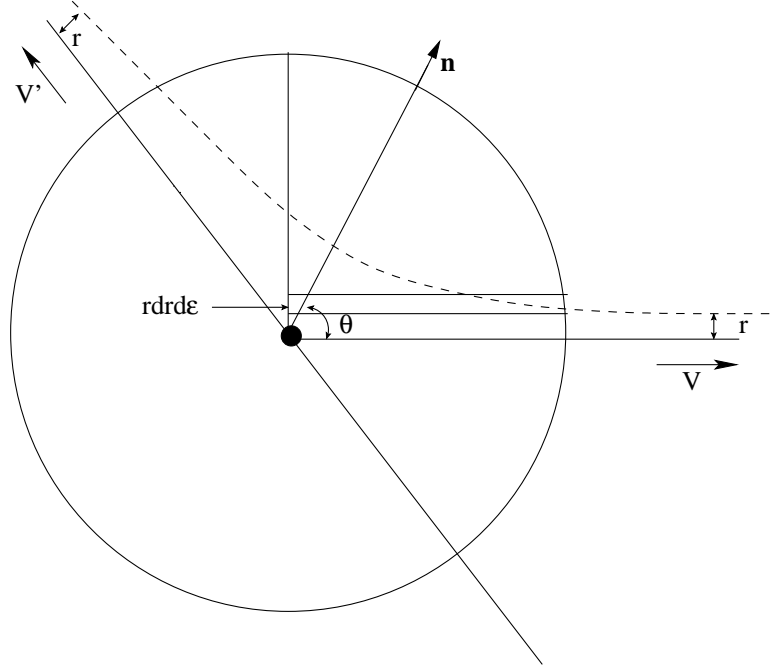 , width=4.0in, height = 3.5in}\caption{The collision coordinates}\label{collision}
\end{center}
\end{figure}

Figure 1 illustrates the coordinates ($\theta, \epsilon$) used for describing the collision. The black dot in the center of the figure refers to the first molecule--the target molecule. The dotted line indicates the trajectory of the second molecule which we call the bullet molecule, with respect to the target molecule. The target molecule is placed at the center where its trajectory comes closest to that of the bullet molecule. We have drawn a sphere around the target molecule and $\mathbf{n}$ is the unit vector in the direction of the point of closest approach of the bullet molecule. The beginning of the trajectory asymptotes in the direction opposite $\mathbf{V}$ and the end of the trajectory asymptotes in the direction opposite $\mathbf{V}^{'}$, which is the relative velocity $\mathbf{\xi}^{'}-\mathbf{\xi}^{*'}$ after the collision. The co-ordinates ($r,\epsilon$) are polar co-ordinates in the plane orthogonal to the plane containing the trajectory of the bullet molecule and the target molecule as shown in the figure. The radial coordinate $r$ is just the impact parameter as shown in the figure. The angular coordinate $\theta$ is the angle between $\mathbf{n}$ and the initial relative velocity $\mathbf{V}$. Thus the unit vector $\mathbf{n}$ is determined by the angular coordinates $\theta$ and $\epsilon$.

Solving Newton's second law for the given central force, we can determine $r$ as a function of $\theta$ and $V$, i.e. if the force is known we know $r(\theta, V)$. The collision kernel $B(\theta, V)$  is defined as
\begin{equation}
B(\theta, V) = V r \frac{\partial r(\theta, V)}{\partial \theta} \quad .
\end{equation}

Finally the velocities of the target and bullet molecule are related to the initial velocities of the target and bullet molecule kinematically through
\begin{eqnarray}
\xi^{'}_{i} &=& \xi_{i} - n_{i}(\mathbf{n \cdot V}) \quad ,\nonumber\\
\xi^{*'}_{i} &=& \xi^{*}_{i} + n_{i}(\mathbf{n\cdot V}) \quad ,
\end{eqnarray}
so that $\mathbf{V^{'}\cdot n} = \mathbf{V\cdot n}$.

This completes our description of the Boltzmann equation. When the molecules interact via an attractive or repulsive central force which is proportional to the fifth inverse power of the distance $\rho$ between the molecules, we say the system is a gas of Maxwellian molecules. The simplification for Maxwellian molecules is that $r$ is independent of $\theta$ which can be seen from the fact that the trajectories of both the target and the bullet molecules lie on the circumference of a circle in the center of mass frame. As a consequence, $B$ is also independent of $\theta$.

To proceed further we need to develop some notation. Let $\phi(\mathbf{\xi})$ be a function of $\xi$. We will call it a collision invariant if
\begin{equation}\label{CInv}
\Phi(\xi, \xi^{*}, \xi^{'}, \xi^{*'}) \equiv \phi(\xi) + \phi(\xi^{*})-\phi(\xi^{'})-\phi(\xi^{*'}) = 0 \quad .
\end{equation}
Clearly there are five collision invariants - $(1, \xi_{i}, \mathbf{\xi}^{2})$ - which we will collectively denote as $\psi_{\alpha}$.

Let us also define, for convenience of notation,
\begin{equation}
\mathcal{J}(f,g) = \frac{J(f,g) + J(g,f)}{2} \quad .
\end{equation}

\textbf{Notation:} We will use the following notation in the rest of this section. Let $A^{(m)}$ and $B^{(n)}$ be two tensors of rank $m$ and $n$ respectively, completely symmetric in all their indices. Then,
\begin{itemize}
\item
$A^{(m)}B^{(n)}$ will denote the symmetric product of the tensors so that it is completely symmetric in all its $m+n$ indices.
\item
$A^{(m)}B^{(n)}_{i}$ will denote a tensor of rank $(m+n)$ where all indices except the $i$-th index in $B^{(n)}$ have been completely symmetrized.
\item
We will use $\nu$ as in $A_{\nu}$ to denote all the $m$ indices in $A$
\item
If $A_{ij}$ and $B_{kl}$ are symmetric second rank tensors, then $(A_{ij}B_{kl}+++++)$ will denote the combination of all the six terms required to make the sum symmetric in its indices $i,j, k$ and $l$.
\end{itemize}
The above notations will hold even when $A$ or $B$ is a tensorial operator containing spatial derivatives.

The hydrodynamic equations can be derived from the Boltzmann equation as follows. Using symmetry one can easily prove that
\begin{equation}\label{J1}
\int\phi(\xi) \mathcal{J}(f,g) d\xi =\frac{1}{4}\int\Phi(\xi,\xi^{*},\xi^{'},\xi^{*'})\mathcal{J}(f,g) d\xi \quad .
\end{equation}

Using (\ref{CInv}) it is clear that if $\phi(\xi)$ is a collision invariant, that is $\phi(\xi)=\psi_{\alpha}(\xi)$, then
\begin{equation}\label{result}
\int \psi_{\alpha}(\xi) \mathcal{J}(f,g) d\xi = 0 \quad .
\end{equation}
A special case of the preceding result gives
\begin{equation}\label{Jprop}
\int \psi_{\alpha}(\xi) J(f,f) d\xi = 0 \quad .
\end{equation}

The Boltzmann equation [on multiplying by $\psi_{\alpha}(\xi)$ and integrating] implies
\begin{equation}\label{hydro}
\frac{\partial \rho_{\alpha}}{\partial t} + \frac{\partial}{\partial x_{i}}\left(\int \xi_{i}\psi_{\alpha}f d\xi\right)=0 \quad ,
\end{equation}
where $\rho_{\alpha}$ are the locally conserved quantities defined by
\begin{equation}\label{locc}
\rho_{\alpha} = \int \psi_{\alpha}f d\xi \quad .
\end{equation}

These equations are equivalent to the hydrodynamic equations (\ref{5mom}) once we make the identifications
[$\rho_{0} = \rho, \ \rho_{i} = \rho u_{i} \ (i=1,2,3),\  \rho_{4} = (3p/2)$].

The next few velocity moments, needed for later reference, are
\begin{eqnarray}
p_{ij} &=& \int (c_{i}c_{j} - RT\delta_{ij}) f d\xi \quad , \nonumber\\
S_{ijk} &=& \int c_{i}c_{j}c_{k} f d\xi \quad ,\\
Q_{ijkl} &=& \int c_{i}c_{j}c_{k}c_{l} f d\xi \quad ,\nonumber
\end{eqnarray}
where $c_{i} = \xi_{i} - u_{i}$.

\subsection*{A.2\ \ The moment equations}
Multiplying both sides of the Boltzmann equation by higher polynomials of $\xi$ and integrating over $\xi$, we find the equations satisfied by the moments $f^{(n)}$'s for $n\geq2$ to be
\begin{equation}\label{momenteq}
\frac{\partial f^{(n)}}{\partial t} + \frac{\partial}{\partial x_{i}}\left(u_{i}f^{(n)}+f_{i}^{(n+1)}\right) +\frac{\partial \mathbf{u}}{\partial x_{i}}f_{i}^{(n)} - \frac{1}{\rho}f^{(n-1)}\frac{\partial f_{i}^{(2)}}{\partial x_{i}} = J^{(n)} \quad ,
\end{equation}
where
\begin{equation}
J^{(n)} = \int \mathbf{c}^{n} B (f^{'}f_{1}^{'}-ff_{1})d\theta d\epsilon d\xi d\xi_{1} \quad ,
\end{equation}
is the $n$-th velocity moment of the collision kernel.

It can be shown that
\begin{equation}
J^{(n)}_{\mu} = \sum_{p,q = 0; p\geq q}^{\infty} B^{(n,p,q)}_{\mu\nu\rho}(\rho, T)f^{(p)}_{\nu}f^{(q)}_{\rho} \quad ,
\end{equation}
with a particular simplification for $B^{(2,2,0)}_{ijkl}$, which can be written as
\begin{equation}\label{Bns}
B^{(2,2,0)}_{ijkl}(\rho, T) = B^{(2)}(\rho, T)\delta_{ik}\delta_{jl} \quad .
\end{equation}

For Maxwellian molecules, there is yet another remarkable simplification that $B^{(n,p,q)}$'s are nonzero only if $p+q =n$. This happens essentially because the collision kernel $B(\theta, V)$ in (\ref{J}) is independent of $\theta$ in this case (for more details please see \cite{Grad}).

We will also denote $f^{(4)}_{ijkl}$ as $Q_{ijkl}$ and its explicit form will be useful.

\subsection*{A.3\ \ Formal Proof of Existence of Conservative Solutions}
We now outline the proof that demonstrates existence of conservative solutions for the Boltzmann equation. The one-particle phase space distribution $f$ will be functionally determined by the hydrodynamic variables and the shear-stress tensor (and their spatial derivatives). It must be emphasized that we proceed $exactly$ along the same lines as used by Enskog in proving the existence of the normal (or purely hydrodynamic solutions) of the Boltzmann equation.

The proof for the existence of normal solutions of the Boltzmann equation \cite{Enskog,Burnett,Chapman,book} (first given in Enskog's thesis) rests on the following theorem due to Hilbert \cite{Hilbert, book}.\\
\textbf{Theorem: }\textit{Consider the following linear integral equation for $g$:}
\begin{equation}
J(f_{0},g) + J(g,f_{0})= \mathcal{K} \quad ,
\end{equation}
\textit{where $J(f_{0}, g)$ is defined through (\ref{J}) and $f_{0}$ is a locally Maxwellian distribution. This equation has a solution if and only if the source term $\mathcal{K}$ is orthogonal to the collision invariants $\psi_{\alpha}$ so that:}
\begin{equation}
\int \psi_{\alpha} \mathcal{K} d\xi = 0 \quad ,
\end{equation}
\textit{provided the potential $U(\rho)$ satisfies the condition that $|U(\rho)|\geq \mathcal{O}(\rho^{-n+1})$ as $\rho\rightarrow 0$ for $n\geq5$. [That is, when the distance ($\rho$) between molecules vanishes, the absolute value of the potential should grow faster than $(1/\rho)^{4}$.] Further the solution is unique up to an additive linear combination of the $\psi_{\alpha}$'s.}\\

This theorem will be important in proving the existence of conservative solutions too, wherein we have to actually solve for the functional dependence on the hydrodynamic variables and the shear-stress tensor. For any conservative solution, we will just need to specify the initial data for the hydrodynamic variables and the shear-stress tensor. The only requirement will be that these initial data are analytic, because the functional dependence of $f$ on the hydrodynamic variables and the shear-stress tensor will involve spatial derivatives of all orders. Clearly all normal solutions are conservative solutions, but not vice versa.

The method of proof can be briefly outlined thus. We will extract a \textit{purely nonhydrodynamic} part from the shear-stress tensor $p_{ij}$, and denote it as $p_{ij}^{(nh)}$. This $p_{ij}^{(nh)}$ will satisfy a simpler equation of motion which schematically reads $(\partial p^{(nh)}/\partial t) = \sum_{n=1}^{\infty} c_{n}(p^{(nh)})^{n}$, involving just a single time derivative [although the initial data for $p_{ij}^{(nh)}$ can have any (analytic) spatial dependence]. The full shear-stress tensor $p_{ij}$ can be solved as a functional of the hydrodynamic variables and the $p_{ij}^{(nh)}$. One can functionally invert this to reinstate $p_{ij}$ as the independent variable in place of $p_{ij}^{(nh)}$ and also determine the equation for $p_{ij}$. In the process we will see that there is an interesting class of nontrivial homogenous conservative solutions, where all the hydrodynamic variables are constants over space and time, while the shear-stress tensor is exactly $p_{ij}^{(nh)}$, which is just a function of time. This class of solutions is thus \textit{purely nonhydrodynamic}, representing equilibration in velocity space.

The proof begins by writing the Boltzmann equation abstractly as
\begin{equation}\label{hilbert}
\mathcal{D} = J(f,f) \quad ,
\end{equation}
where
\begin{equation}\label{D}
\mathcal{D} = \frac{\partial f}{\partial t} + \mathbf{\xi} \cdot \frac{\partial f}{\partial \mathbf{x}} \quad ,
\end{equation}
and $J(f,f)$ is as defined through (\ref{J}).

For a conservative solution, $f$ is a functional of the nonhydrodynamic shear-stress tensor, $p_{ij}^{(nh)}(\mathbf{x},t)$ and the five hydrodynamic variables, namely $u_{i}(\mathbf{x},t)$, $\rho(\mathbf{x},t)$ and $T(\mathbf{x},t)$. We expand $f$ in two formal expansion parameters $\epsilon$ and $\delta$ such that
\begin{equation}\label{suboff}
f = \sum_{n=0}^{\infty}\sum_{m=0}^{\infty}\epsilon^{n}\delta^{m}f_{(m,n)} \quad .
\end{equation}

The physical meanings of the expansion parameters will soon be made precise. For the moment, if the reader so pleases, she can think of $\epsilon$ as a hydrodynamic and $\delta$ as a nonhydrodynamic expansion parameter.
Following Enskog, we will also expand the time derivative in powers of $\epsilon$ and $\delta$ as :
\begin{equation}
\frac{\partial}{\partial t} = \sum_{n=1}^{\infty}\sum_{m=0}^{\infty}\epsilon^{n}\delta^{m}\frac{\partial^{(n,m)}}{\partial t} \quad .
\end{equation}
The above expansion of the time derivative might seem a little strange, but it will be necessary for us precisely for the same reason it was necessary for Enskog - the solutions of the equations of motion of hydrodynamic variables and $p_{ij}^{(nh)}$ cannot be expanded analytically in $\epsilon $ and $\delta$, though their equations of motion could be through the subdivision of the
partial time derivative. The proof will actually rely on the subdivision of the equations of motion just as in Enskog's purely hydrodynamic normal solutions and will not require the solutions to have analytic expansions \cite{footnote8}. 

This automatically results in a similar expansion for $\mathcal{D}$, such that
\begin{itemize}
\item
For $n \geq 1$ and for all $m$
\begin{equation}\label{subofd1}
\mathcal{D}^{(n,m)} \equiv  \sum_{k=1}^{n}\sum_{l=0}^{m}\frac{\partial^{(k,l)} f_{(n-k,m-l)}}{\partial t} + \mathbf{\xi} \cdot \frac{\partial f_{(n-1,m)}}{\partial \mathbf{x}} \quad. \quad (n\geq 1; m=0,1,2,...)
\end{equation}
\item
For $n=m=0$,
\begin{equation}\label{subofd2}
\mathcal{D}^{(0,0)} = 0 \quad .
\end{equation}
\end{itemize}

With the assumption that $f$ is a functional of the hydrodynamic variables and $p_{ij}^{(nh)}$,  the time derivative acts on $f$ schematically as
\begin{eqnarray}\label{timed}
\frac{\partial f}{\partial t} &=& \sum_{k=0}^{\infty}\frac{\partial f}{\partial (\nabla^{k} \rho)}\frac{\partial (\nabla^{k}\rho)}{\partial t} +\sum_{k=0}^{\infty}\frac{\partial f}{\partial (\nabla^{k} u_{i})}\frac{\partial (\nabla^{k} u_{i})}{\partial t} \nonumber\\
&& + \sum_{k=0}^{\infty}\frac{\partial f}{\partial (\nabla^{k} T)}\frac{\partial (\nabla^{k} T)}{\partial t} + \sum_{k=0}^{\infty}\frac{\partial f}{\partial (\nabla^{k}p_{ij}^{(nh)})}\frac{\partial (\nabla^{k}p_{ij}^{(nh)})}{\partial t} \quad .
\end{eqnarray}
Above, $\nabla^{k}$ schematically denotes $k$-th order spatial derivatives. Any time derivative acting on a hydrodynamic variable can be replaced by a functional of the hydrodynamic variables and the nonhydrodynamic shear-stress tensor by using the hydrodynamic equations of motion. These functional forms have a systematic derivative expansion in terms of the number of spatial derivatives present and contain only spatial derivatives and no time derivatives. So the expansion of the time derivative in $\epsilon$ is actually a derivative expansion, where the expansion parameter $\epsilon$ is the ratio of the typical length scale of spatial variation of $f$ and the mean-free path. This naturally ``explains'' (\ref{subofd1}).

On the other hand, it will be seen that the time derivative of the nonhydrodynamic shear-stress tensor can be replaced, using its equation of motion, by an infinite series of polynomials of the nonhydrodynamic shear-stress tensor. Thus the expansion of the time derivative in $\delta$ as in (\ref{subofd2}); but we expand the solution of the equation of motion as an amplitude expansion with the expansion parameter $\delta$ identified as the ratio of the typical amplitude of the nonhydrodynamic shear-stress tensor with the pressure in final equilibrium. For the moment, these are just claims, to be borne out by an appropriate definition of the expansion of $f$ and the time derivative.

\begin{flushleft}
\textbf{A.3.1 Subdivisions in terms of $\epsilon$ and $\delta$}
\end{flushleft}
We outline here the expansion of the various quantities in the Boltzmann equation and the full Boltzmann equation itself in terms of the two expansion parameters $\epsilon$ and $\delta$ and thereby arrive at various constraints that must be satisfied by these expansions. Our proof eventually will involve recursion while expanding the full Boltzmann equation in these expansion parameters.
\begin{enumerate}
\item
In close analogy with Enskog's original subdivision of $f$, we impose some further properties on the subdivision of $f$.
\begin{itemize}
\item
First we require, as in the case of normal solutions of Enskog and Chapman, that the hydrodynamic variables are unexpanded in $\epsilon$ and $\delta$ and therefore are exactly the same as in the zeroth-order solution $f_{(0,0)}$, which will turn out to be locally Maxwellian. This is required because solutions of the hydrodynamic equations cannot be expanded analytically in these expansion parameters though the hydrodynamic equations themselves could be, as mentioned above. Therefore we should have
\begin{equation}\label{conf}
\int \psi_{\alpha}f_{(n,m)}d\xi =0; \qquad (n+m \geq 1;\alpha = 0,1,2,3,4)
\end{equation}
where $\psi_{\alpha}$ are the collision invariants $(1,\xi_{i}, \mathbf{\xi}^{2})$.
It follows that
\begin{equation}\label{zorder}
\rho_{\alpha} = \int \psi_{\alpha}f_{(0,0)} d\xi \quad .
\end{equation}
$\rho_{\alpha}$ are the locally conserved quantities defined through (\ref{locc}). We may recall that these are just some combinations of the hydrodynamic variables. 
\item
We also require that the purely nonhydrodynamic part of the shear-stress tensor, $p_{ij}^{(nh)}$ has no expansion in $\epsilon$ and $\delta$, analogous to the hydrodynamic variables. Being purely nonhydrodynamic, it determines $f_{(0,m)}$ for all $m$, i.e. the part of $f$ which is zeroth order in $\epsilon$, but contains all orders of $\delta$ in the conservative solutions. Since it vanishes at equilibrium, it is of first order in $\delta$ and is given exactly by $f_{(0,1)}$. More explicitly, for $m\geq2$ and $n=0$, we should have
\begin{equation}
\int (c_{i}c_{j} -RT\delta_{ij})f_{(0,m)}d\xi =0 \quad (m\geq2) \quad ,
\end{equation}
so that
\begin{equation}\label{m=1}
p_{ij}^{(nh)} = \int (c_{i}c_{j} -RT\delta_{ij})f_{(0,1)}d\xi \quad .
\end{equation}
\end{itemize}
\item
The subdivision of the time derivative is defined next. Following Enskog, we impose on the time-derivative the condition that
\begin{eqnarray}\label{suboft}
\frac{\partial^{(0,m)}\rho^{\alpha}}{\partial t} &=& 0 \quad , \nonumber\\
\int\mathcal{D}^{(n,m)}\psi_{\alpha}d\xi &=& 0; \quad (n\geq1,m=0,1,2,...) \quad.
\end{eqnarray}
Using (\ref{conf}), the second equation above can be simplified to
\begin{equation}\label{suboft2}
\frac{\partial^{(n,m)}\rho^{\alpha}}{\partial t} + \frac{\partial}{\partial x_{i}}\left(\int \xi_{i}\psi_{\alpha}f_{(n-1,m)}d\xi\right) = 0 \quad (n \geq 1,m=0,1,2,..) \quad .
\end{equation}
Since the $\rho^{\alpha}$ are a redefinition of the hydrodynamic variables, this above condition amounts to expanding the hydrodynamic equations in a particular way. From this expansion we know how each subdivision of the time derivative acts on the (unexpanded) hydrodynamic variables. It is clear from (\ref{timed}) that if we now specify how the subdivisions of the time derivative act on $p_{ij}^{(nh)}$, we have defined the time derivative. Indeed, we have to solve for the action of the time-derivative because specifying this amounts to proving the existence of conservative solutions \cite{footnote9}. 

\item
The next thing is to note that the full shear-stress tensor $p_{ij}$ (just like any other higher moment) has an expansion in both $\epsilon$ and $\delta$. If we denote $\delta p_{ij} = p_{ij} - p_{ij}^{(nh)}$, then for $n\geq 1$
\begin{equation}
\delta p_{ij}^{(n,m)} = \int \left(c_{i}c_{j} - RT\delta_{ij}\right) \left(f_{(n,m)}-f_{(0,1)}\right) d\xi \quad ,
\end{equation}
need not vanish.
The expansion of $\delta p_{ij}^{(n,m)}$ as a functional of the hydrodynamic variables and $p_{ij}^{(nh)}$ in $\epsilon$ is the derivative expansion, with the power of $\epsilon$ essentially counting the number of spatial derivatives (which act both on hydrodynamic variables and the nonhydrodynamic shear-stress tensor). The expansion in $\delta$ is the ``amplitude'' expansion in terms of $p_{ij}^{(nh)}$, which we may recall is first order in $\delta$.
\item
On the basis of the above subdivisions one can now expand both sides of (\ref{hilbert}) and equate the terms of the same order on both sides. This enables us to write down the following set of equations that $J(f,f)$ must satisfy for different values of $(n,m)$.
\begin{itemize}
\item
For $n=m=0$, substituting (\ref{subofd1}), (\ref{subofd2}) and (\ref{suboff}) in (\ref{hilbert}) we get
\begin{equation}\label{subofb1}
J(f_{(0,0)}, f_{(0,0)}) = 0 \quad ,
\end{equation}
so that $f_{(0,0)}$ has to be a locally Maxwellian distribution.
\item
Using the above fact, for $n=0$ and $m \geq 1$, we get
\begin{eqnarray}\label{subofb2}
J(f_{(0,0)},f_{(0,m)}) + J(f_{(0,m)},f_{(0,0)}) - \frac{\partial^{(0,0)}}{\partial t}f_{(0,m)} \nonumber\\
= \sum_{l=1}^{m}\frac{\partial^{(0,l)}}{\partial t}f_{(0,m-l)} -S_{(0,m)}; \quad (m\geq1) \quad .
\end{eqnarray}
\item
Finally, for $n\geq 1$ and for all $m$
\begin{eqnarray}\label{subofb3}
J(f_{(0,0)}, f_{(n,m)}) + J(f_{(n,m)}, f_{(0,0)}) = \qquad \qquad \nonumber\\
\sum_{k=1}^{n}\sum_{l=0}^{m}\frac{\partial^{(k,l)} f_{(n-k,m-l)}}{\partial t} + \mathbf{\xi} \cdot \frac{\partial f_{(n-1,m)}}{\partial \mathbf{x}} - S_{(n,m)};  \quad (n\geq1).
\end{eqnarray}
\item
The $S_{(n,m)}$ are given by, for $(n+m \geq 2)$
\begin{eqnarray}
S_{(n,m)} = \sum_{k=1}^{n-1}\sum_{l=1}^{m-1}J(f_{(k,l)},f_{(n-k,m-k)}) \nonumber\\
+ \sum_{k=1}^{n-1}J(f_{(k,0)},f_{(n-k,m)})+ \sum_{l=1}^{m-1}J(f_{(0,l)},f_{(n,m-k)})\\\nonumber +\sum_{k=1}^{n-1}J(f_{(k,m)},f_{(n-k,0)})+ \sum_{l=1}^{m-1}J(f_{(n,l)},f_{(0,m-k)})\\
+ J(f_{(n,0)},f_{(0,m)}) + J(f_{(0,m)},f_{(n,0)});  \quad (n+m)\geq 2 \nonumber
\end{eqnarray}
and
\begin{equation}
S_{(0,1)} = S_{(1,0)} = 0 \quad .
\end{equation}
\end{itemize}
\end{enumerate}

\begin{flushleft}
\textbf{A.3.2 A recursive proof}
\end{flushleft}
With all of the above, we will now prove the existence and uniqueness of conservative solutions recursively. Recall that the key idea in this proof is to understand how the time derivative operator $\frac{\partial}{\partial t}$ acts on the hydrodynamic variables and the nonhydrodynamic part of the shear-stress tensor, $p^{(nh)}_{ij}$. We already know the action of this operator on the hydrodynamic variables from Eqs.(\ref{suboft}) and (\ref{suboft2}). Now we will solve for the action of this operator on $p_{ij}^{(nh)}$. The action of the time derivative, when expanded in $\epsilon$ and $\delta$, can be understood by analyzing the subdivisions of the Boltzmann equation given by Eqs. (\ref{subofb1}), (\ref{subofb2}) and (\ref{subofb3}).

\begin{enumerate}
\item
It is clear from (\ref{subofb1}) that at the zeroth-order in $m$ and $n$, $f_{(0,0)}$ is a locally Maxwellian distribution which is uniquely fixed by the choice of the five hydrodynamic variables (\ref{zorder}) and hence can uniquely be specified as
\begin{equation}
f_{(0,0)}=\frac{\rho}{(2\pi RT)^{\frac{3}{2}}}\exp{\left(-\frac{\mathbf{c}^{2}}{2RT}\right)} \quad .
\end{equation}
\item
Next let us consider (\ref{subofb2}). The usual trick here is to rewrite $f_{(0,m)}$ as $f_{(0,0)}h_{(0,m)}$. The advantage is that since $f_{(0,0)}$ contains hydrodynamic variables only,
\begin{equation}
\frac{\partial^{(0,m)}}{\partial t}f_{(0,0)} = 0 \quad .
\end{equation}

Therefore (\ref{subofb2}) can be rewritten as
\begin{align}\label{rsubofb2}
J(f_{(0,0)},f_{(0,0)}h_{(0,m)}) + J(f_{(0,0)}h_{(0,m)},f_{(0,0)}) - f_{(0,0)}\frac{\partial^{(0,0)}}{\partial t}h_{(0,m)} \qquad \\\nonumber = f_{(0,0)}\sum_{l=1}^{m}\frac{\partial^{(0,l)}}{\partial t}h_{(0,m-l)} - \sum_{l=1}^{m-1}J(f_{(0,0)}h_{(0,l)},f_{(0,0)}h_{(0,m-l)});  \qquad (m \geq 2).
\end{align}
Now we analyze  (\ref{rsubofb2}) for $m=1$ and $m=2$.\\
\begin{itemize}
\item \textbf{m=1}:\\

For $m=1$, (\ref{rsubofb2}) reduces to
\begin{align}
J(f_{(0,0)},f_{(0,0)}h_{(0,1)}) + J(f_{(0,0)}h_{(0,1)},f_{(0,0)})\nonumber\\  = f_{(0,0)}\frac{\partial^{(0,0)}}{\partial t}h_{(0,1)} \quad ,
\end{align}
while it follows from (\ref{m=1}) that
\begin{equation}
h_{(0,1)} = \frac{1}{2!}\frac{p_{ij}^{(nh)}(\mathbf{x},t)}{pRT}(c_{i}c_{j}-RT\delta_{ij}) \quad .
\end{equation}

These two equations imply that
\begin{equation}\label{firsteop}
\frac{\partial^{(0,0)}}{\partial t}p_{ij}^{(nh)} = B^{(2)}(\rho,T)p_{ij} \quad ,
\end{equation}

where $B^{(2)}$ has been defined in (\ref{Bns}) \cite{footnote10}. \\

\item \textbf{m=2}:\\
At the second order, (\ref{rsubofb2}) implies
\begin{align}\label{seceop1}
J(f_{(0,0)},f_{(0,0)}h_{(0,2)}) + J(f_{(0,0)}h_{(0,2)},f_{(0,0)}) \qquad \nonumber\\
= f_{(0,0)}\frac{\partial^{(0,0)}}{\partial t}h_{(0,2)} + f_{(0,0)}\frac{\partial^{(0,1)}}{\partial t}h_{(0,1)} - J(f_{(0,0)}h_{(0,1)},f_{(0,0)}h_{(0,1)}) \quad .
\end{align}

We then need to solve for two things, $h_{(0,2)}$ and the operator $\left(\partial^{(0,1)}/\partial t\right)$. To do this we first write $h_{(0,2)}$ as

\begin{align}
h_{(0,2)} = \frac{1}{3!}\frac{S_{ijk}^{(0,2)}}{p(RT)^2}\Big(c_{i}c_{j}c_{k}-RT(c_{i}\delta_{jk}++)\Big) \qquad \\\nonumber
+ \frac{1}{4!(RT)^2}\left[\frac{Q_{ijkl}^{(0,2)}}{p(RT)}-\left(\frac{p_{ij}^{(nh)}}{p}\delta_{kl}+++++\right)-\left(\delta_{ij}\delta_{kl}++\right)\right]\\\nonumber
\times \Big[c_{i}c_{j}c_{k}c_{l}-RT(c_{i}c_{j}\delta_{kl}+++++)+(RT)^{2}(\delta_{ij}\delta_{kl}++)\Big] \quad .
\end{align}

The idea behind guessing this form is to expand $h_{(0,2)}$ in two higher order Hermite polynomials of $\mathbf{c}$'s and reexpressing the Hermite coefficients through the ordinary moments. This method of expansion is due to Grad \cite{Grad}. For the moment we can just take it as the most general possible form of $h_{(0,2)}$, since if higher Hermite polynomials are included here, the coefficients would have vanished. It also turns out that $S_{ijk}^{(0,2)}$ vanishes. Similarly all the other higher odd moments  vanish, \textit{so far as their purely nonhydrodynamic parts (or expansion in $m$ for $n=0$) is concerned}. Obviously this does not mean that these higher odd moments have no dependence on $p_{ij}^{(nh)}$. For $n>0$ there is indeed a nonvanishing expansion in $m$ for these moments.  We can now compare the coefficients of Hermite polynomials on both sides of our Eq. (\ref{seceop1}). For Maxwellian molecules (thus determining the form of $J$) we have

\begin{eqnarray}\label{seceop}
\frac{\partial^{(0,1)}}{\partial t}p_{ij}^{(nh)} &=& B^{(2,2,2)}_{ijklmn}(\rho , T)p_{kl}^{(nh)}p_{mn}^{(nh)} + B^{(2,4,0)}_{ijklmn}(\rho, T)Q_{klmn}^{(0,2)} \quad ,\\\nonumber
\frac{\partial^{(0,0)}}{\partial t}Q_{ijkl}^{(0,2)} &=& B^{(4,4,0)}_{ijklmnpq}(\rho, T)Q_{mnpq}^{(0,2)} +B^{(4,2,2)}_{ijklmnpq}(\rho,T)p_{mn}^{(nh)}p_{pq}^{(nh)} \quad .
\end{eqnarray}

Since we know the action of $(\partial^{(0,0)}/\partial t)$ on $p_{ij}^{(nh)}$ and the hydrodynamic variables, we can solve for $Q_{ijkl}^{(0,2)}$ as a functional of $p_{ij}^{(nh)}$ and the hydrodynamic variables; the solution turns out to be

\begin{equation}
Q_{klmn}^{(0,2)} = X_{klmnpqrs}p_{pq}^{(nh)}p_{rs}^{(nh)} \quad ,
\end{equation}

where $X_{klmnpqrs}$ satisfies the equation \cite{footnote11}

\begin{equation}
2 B^{(2)}X_{klmnpqrs} = B^{(4,4,0)}_{klmnijtu}X_{ijtupqrs} + B^{(4,2,2)}_{klmnpqrs} \quad .
\end{equation}

This in turn provides the solution for the operator $\left(\partial^{(0,1)}/\partial t\right)$:

\begin{equation}
\frac{\partial^{(0,1)}}{\partial t}p_{ij}^{(nh)} = B^{(2,2,2)}_{ijklmn}p_{kl}^{(nh)}p_{mn}^{(nh)} +  B^{(2,4,0)}_{ijklmn}X_{klmnpqrs}p_{pq}^{(nh)}p_{rs}^{(nh)} \quad .
\end{equation}

The equation above shows the action of the operator on $p_{ij}^{(nh)}$; we already know how it acts on the hydrodynamic variables. This implies that we have solved for this operator at this order. Note that the solution for the operator corroborates the intuitive understanding that this operator is the next order in amplitude expansion. Another important point is that the solution of the operator is not independent of the solution for $Q_{ijkl}^{(0,2)}$ and is just given by the logic of our expansion once $Q_{ijkl}^{(0,2)}$ has been solved as a functional of $p_{ij}^{(nh)}$. This feature is the same for all the higher terms in the expansion of the time derivative operator as well.

For non-Maxwellian molecules things are a bit complicated because the equation for $Q_{ijkl}$ in (\ref{seceop}) also contains a term linear in $p_{ij}^{(nh)}$, so that now

\begin{eqnarray}
\frac{\partial^{(0,1)}}{\partial t}p_{ij}^{(nh)} &=& \delta B^{(2)}p_{ij}^{(nh)}+B^{(2,2,2)}_{ijklmn}p_{kl}^{(nh)}p_{mn}^{(nh)} \nonumber\\
&&+ B^{(2,4,0)}_{ijklmn}X_{klmnpqrs}p_{pq}^{(nh)}p_{rs}^{(nh)} \quad .
\end{eqnarray}

However this feature also appears in the usual derivative expansion (the expansion in $\epsilon$) of the time-derivative. Despite appearance, $\delta B^{(2)}p_{ij}^{(nh)}$ is a small quantity as $(\delta B^{(2)}/B^{(2)})$ is a pure number which is smaller than unity (for a proof of this and also for the statement of convergence of such corrections in the context of normal solutions, please see \cite{Burnett,Chapman}). This result can be translated here, as the normal solutions are just special cases of our conservative solutions and at a sufficiently late time our solutions will be just appropriate normal solutions \cite{footnote12}.

This is indeed remarkable considering that we have no parametric suppression here. Formally however, aside from the convergence problem, there is no obstruction because $\delta$ is just a formal parameter and is only intuitively connected to the amplitude expansion.\\
\item \textbf{Higher m}:\\
We can proceed in the same way to the next order in $m$ when $n$ is zero. At every stage we have to deal with $f_{(0,m)}$ which we may write as $f_{(0,0)}h_{(0,m)}$ and further expand $h_{(0,m)}$ in a series containing up to $m$-th order Hermite polynomial in $c$'s. We have to solve for the coefficients of these Hermite polynomials, which depend on $x$ only, and this leads to the definition of the $m$-th subdivision of the time derivative operator in the $\delta$ expansion when the $\epsilon$ expansion is at the zeroth order. The equation for evolution of $p_{ij}^{(nh)}(\mathbf{x},t)$ thus finally involves only a single time derivative which we have expanded in $\delta$. This is highly nonlinear, involving an infinite series of $p_{ij}^{(nh)}$. The presence of just a single time derivative in the equation of motion for  $p_{ij}^{(nh)}(\mathbf{x},t)$ makes it essentially an ordinary differential equation in one variable and so for any initial data existence and uniqueness of solution is guaranteed.

We note that we can consistently truncate our solution at $n=0$ so that there is no expansion of $f$ in $\epsilon$, provided all the hydrodynamic variables are constants over both space and time and $p_{ij}^{(nh)}$ is constant over space but a function of time. This  gives us the simplest class of conservative solutions which is homogenous in space; the Boltzmann equation becomes equivalent to an ordinary differential equation involving a single time derivative for $p_{ij}$. Physically this solution corresponds to the most general conservative solution which is homogenous in space, but generically far away from equilibrium in the velocity space (so that the velocity distribution is far from being Maxwellian).
\end{itemize}
\item
The next task is to prove the existence of solutions for the recursive series of equations in (\ref{subofb3}). To see if solutions will exist we need to employ Hilbert's theorem. $S_{(n,m)}$ contains either pairs of the form $J(f_{(p,q)},f_{(r,s)})+J(f_{(p,q)},f_{(r,s)})$ or just $J(f_{(l,l)},f_{(l,l)})$. So when the collision invariants are integrated with $S_{(n,m)}$, as in $\int \psi_{\alpha} S_{(n,m)} d\xi$, the integrals vanish as a consequence of (\ref{result}). Therefore, the existence of the solution to $f_{(n,m)}$ follows from Hilbert's theorem as a consequence of (\ref{suboft}). The solution is unique because the condition (\ref{conf}) fixes the arbitrariness of the dependence of $f_{(n,m)}$ on the collision invariants $\psi_{\alpha}$. The details for $n \geq 1$, are thus, exactly the same as in the case of normal solutions. The action of $(\partial^{(n,m)}/\partial t)$ on $p_{ij}^{(nh)}$ is also determined as soon as the functional dependence of $\delta p_{ij}$ and the relevant higher moments on $p_{ij}^{(nh)}$ and the hydrodynamic variables are determined.

The explicit calculations become extremely complex even when, say $n=2, m=0$ or $n=1, m=1$. We give some explicit results for the first few terms in the expansion for $\delta p_{ij}$ as 

\begin{eqnarray}\label{fluckin}
\delta p_{ij}^{(1,0)} &=&  \frac{p}{B^{(2)}} (\frac{\partial u_{m}}{\partial x_{n}}+\frac{\partial u_{n}}{\partial x_{m}}-\frac{2}{3}\delta_{mn}\frac{\partial u_{r}}{\partial x_{r}}) \quad ,\\\nonumber
\delta p_{ij}^{(1,1)} &=&  \frac{1}{B^{(2)}}\left(\frac{\partial }{\partial x_{r}}(u_{r} p_{ij}^{(nh)}) +\frac{\partial u_{j}}{\partial x_{r}} p_{ir}^{(nh)} + \frac{\partial u_{i}}{\partial x_{r}} p_{jr}^{(nh)} -\frac{2}{3}\delta_{ij}p_{rs}^{(nh)}\frac{\partial u_{r}}{\partial x_{s}}\right) \\\nonumber
&&-\frac{2 pB^{(2,2,2)}_{ijklmn}}{(B^{(2)})^{2}}p_{kl}^{(nh)}(\frac{\partial u_{m}}{\partial x_{n}}+\frac{\partial u_{n}}{\partial x_{m}}-\frac{2}{3}\delta_{mn}\frac{\partial u_{r}}{\partial x_{r}}) \quad .
\end{eqnarray}

It is clear that the terms in the expansion involve spatial derivatives of both the hydrodynamic variables and $p_{ij}^{(nh)}$. From the expression for $\delta p_{ij}^{(1,0)}$ one can determine the shear viscosity $\eta$ which is of course the same as in the purely hydrodynamic normal solutions, so that
\begin{equation}
\eta \approx \frac{p}{B^{(2)}}(\rho, T) \quad .
\end{equation}

We also give some terms in the expansion for the heat flow vector
\begin{eqnarray}
S_{i}^{(1,0)} &=& \frac{15pR}{2B^{(2)}}\frac{\partial T}{\partial x_{i}} \quad ,\\\nonumber
S_{i}^{(1,1)} &=& \frac{3}{2B^{(2)}}\left(2RT\frac{\partial p_{ir}^{(nh)}}{\partial x_{r}}+ 7R p_{ir}^{(nh)}\frac{\partial T}{\partial x_{r}} - \frac{2p_{ir}^{(nh)}}{\rho}\frac{\partial p}{\partial x_{r}}\right) \quad ,
\end{eqnarray}

It is clear that the heat conductivity $\chi$ is also the same as in purely hydrodynamic normal solutions so that
\begin{equation}
\chi \approx \frac{15R}{2}\frac{p}{B^{(2)}}(\rho, T) \approx \frac{15 R}{2}\eta \quad .
\end{equation}
Corrections to the above relation appear in the higher order for non-Maxwellian molecules but again these are the same as in the case of normal solutions.
\end{enumerate}

This completes our proof for the existence of conservative solutions for the nonrelativistic Boltzmann equation.
As mentioned before, we can now reinstate $p_{ij}$ as the independent variable. Our independent variables satisfy the following equations of motion

\begin{eqnarray}\label{9momApp}
\frac{\partial \rho}{\partial t} + \frac{\partial}{\partial x_{r}}(\rho u_{r}) &=& 0 \quad ,\nonumber\\
\frac{\partial u_{i}}{\partial t} + u_{r} \frac{\partial u_{i}}{\partial x_{r}} + \frac{1}{\rho}\frac{\partial (p\delta_{ir}+p_{ir})}{\partial x_{r}} &=& 0 \quad ,\nonumber\\
\frac{\partial p}{\partial t} + \frac{\partial}{\partial x_{r}}(u_{r}p) +\frac{2}{3}(p\delta_{ir}+p_{ir})\frac{\partial u_{i}}{\partial x_{r}} +\frac{1}{3}\frac{\partial S_{r}}{\partial x_{r}} &=& 0  \quad ,\\
\frac{\partial p_{ij}}{\partial t} +\frac{\partial}{\partial x_{r}}(u_{r}p_{ij}) +\frac{\partial S_{ijr}}{\partial x_{r}} -\frac{1}{3}\delta_{ij}\frac{\partial S_{r}}{\partial x_{r}} \nonumber\\
+\frac{\partial u_{j}}{\partial x_{r}} p_{ir}+ \frac{\partial u_{i}}{\partial x_{r}} p_{jr} -\frac{2}{3}\delta_{ij}p_{rs}\frac{\partial u_{r}}{\partial x_{s}} \nonumber\\
+p(\frac{\partial u_{i}}{\partial x_{j}}+\frac{\partial u_{j}}{\partial x_{i}}-\frac{2}{3}\delta_{ij}\frac{\partial u_{r}}{\partial x_{r}}) &=& \sum_{p,q = 0,p\geq q; (p,q) \neq (2,0))}^{\infty} B^{(2,p,q)}_{ij\nu\rho}(\rho, T)f^{(p)}_{\nu}f^{(q)}_{\rho}\nonumber\\
&& +B^{(2)}(\rho, T)p_{ij}\nonumber \quad .
\end{eqnarray}

The first three equations are just the hydrodynamic equations, while the equation for $p_{ij}$ can be obtained from (\ref{momenteq}).

The crucial point of this proof is that we have now solved for all higher moments $f^{(n)}_{\nu}$'s for $n\geq3$
(which includes, of course, $S_{ijk}$ and thus $S_{i}$) as functionals of our ten variables ($\rho, u_{i}, p, p_{ij}$) with  $T = p/(R\rho)$. Any solution of these ten equations of motion can be uniquely lifted to a full solution of the Boltzmann equation as all the higher moments are dependent on these ten variables through a unique functional form. Also, some solutions for $p_{ij}$ in the last of our system of equations are purely hydrodynamic and these constitute the normal solutions \cite{footnote13}. 

\providecommand{\href}[2]{#2}\begingroup\raggedright


\begin{thebibliography}{06}
\bibitem{footnote1}
The universality can happen even when we have nonconformal versions of AdS/CFT or consider specific $1/N$ and finite coupling corrections in the conformal case \cite{other}.
\bibitem{footnote2}
An example of interest is the formation of quark-gluon plasma through collisions of gold nuclei, which happens spontaneously without the direct influence of environment. Strictly speaking, such systems are never isolated since they may lose energy through low energy radiation and we need to average out the boundary conditions determining the radiation escaping to infinity appropriately, so that the final equilibrium state is specified.
\bibitem{footnote}
If we refine the kinetic description beyond the Boltzmann equation, we need to refine this equation of state which holds for ideal gases. The gaseous equation of state assumes that the potential energy density is negligible compared to the kinetic energy density which could be true only if the number density of particles is sufficiently small.
\bibitem{footnote4}
Classically this is just the typical time it takes the trajectories of the molecules to straighten out after collision; a good estimate of this is $r/c_{s}$, where $r$ is the range of the force and $c_{s}$ is the thermal speed (the average root mean square velocity of the particles).
\bibitem{footnote6}
The proof follows along exactly the same lines as shown in Appendix. The Boltzmann equation still takes the same form as in (\ref{be}). The semiclassical form of $J$ then readily follows from (\ref{Jprop}), which in turn follows from (\ref{J1}) and (\ref{result}), all of which are true for the semiclassical form of $J$ too. The hydrodynamic equations will take the same form as before and the shear-stress tensor, $p_{ij}$ and the heat flow vector $S_{i}$ can be defined as before.
\bibitem{footnote7}
The explicit solutions in the recursive expansion series will be more complicated now. In the nonrelativistic proof, one uses the Hermite polynomials which can no longer be conveniently employed.
\bibitem{footnote14}
Even in the purely hydrodynamic context of "tubewise black-brane solutions," the Israel-Stewart-Muller formalism is not valid. This has been discussed with references later in the text.
\bibitem{footnote8}
There is an analog of this in the fluid/gravity correspondence too. The existence of solutions in gravity dual to hydrodynamic configurations in the boundary which could be analytically expanded depends on the derivative expansion of the hydrodynamic equations and is independent of the fact that the solutions of the hydrodynamic equations themselves have no analytic expansion in the derivative expansion parameter.
\bibitem{footnote9}
Even for the hydrodynamic variables, the action of the subdivisions of the time derivative on them could have been treated as unknowns. But we have chosen the logically equivalent path of declaring them beforehand from our experience with the case of normal solutions.
\bibitem{footnote10}
Note that (\ref{firsteop}) is consistent with the amplitude expansion of the time-derivative in $\delta$. The time-derivative expansion in $\delta$ should consistently start at the zeroth order as both sides of the equation contain one $p_{ij}^{(nh)}$.
\bibitem{footnote11}
The solution for $X_{(klmn)(pqrs)}$ regarded as an $81\times81$ matrix is $(2B^{(2)}\delta_{(klmn)(ijtu)}-B^{(4,4,0)}_{(klmn)(ijtu)})^{-1}B^{(4,2,2)}_{(ijtu)(pqrs)}$ where $\delta_{(klmn)(ijtu)}$ is defined as a $81\times81$ matrix whose entries are $1$ if $k=i,\ l=j, \ m=t, \ n=u$ and zero otherwise. It is quite evident that when $(2B^{(2)}\delta_{(klmn)(ijtu)}-B^{(4,4,0)}_{(klmn)(ijtu)})$ fails to be invertible, there is a singularity in our solution and in fact this may happen when $\rho$ and $T$ takes appropriate values. Such singularities also appeared in Born and Green's normal solutions of BBGKY heirarchy and was interpreted as describing local nucleation of the solid phase. In our case too, the singularities of the conservative solutions may signal local condensation of the liquid phase.
\bibitem{footnote12}
We note that we can simply borrow Burnett's results here because all the hydrodynamic transport coefficients are not independent of the nonhydrodynamic parameters like the relaxation time $1/B^{(2)}$, for instance viscosity $\eta$ is at leading order $(p/B^{(2)})$. Since $p$ has no expansion, convergence of the viscosity implies convergence of $B^{(2)}$ as any generic ``conservative'' solution will be approximated by an appropriate normal solution at sufficiently late times.
\bibitem{footnote13}
This can be readily seen as follows. If we assume that $p_{ij}$ is functionally dependent on the hydrodynamic variables, from the equation for its evolution, it is clear that at the first order in the derivative expansion $p_{ij} = \eta(\frac{\partial u_{i}}{\partial x_{j}}+\frac{\partial u_{j}}{\partial x_{i}}-\frac{2}{3}\delta_{ij}\frac{\partial u_{r}}{\partial x_{r}})$, where $\eta = \frac{p}{B^{(2)}}$. Now we can substitute this in place of $p_{ij}$ in the equation of evolution for $p_{ij}$ to get the second order correction and so on. In the substitution, the time derivative acting on the hydrodynamic variables can be replaced by spatial derivatives by using the hydrodynamic equations of motion.
\end{thebibliography}
\end{document}